\documentclass[11pt]{article}
\usepackage{amssymb,latexsym,amsmath,amsthm,graphicx,amscd}
\usepackage{upgreek} 

\newtheorem{theorem}{Theorem}
\makeatletter \@addtoreset{figure}{section}
\def\fps@figure{h, t}
\@addtoreset{table}{bsection}
\def\thetable{\thesection.\@arabic\c@table}
\def\fps@table{h, t}

\newcommand{\bear}{\begin{array}}
\newcommand{\eear}{\end{array}}

\newtheorem{lemma}[theorem]{Lemma}
\newtheorem{proposition}[theorem]{Proposition}

\newtheorem{cor}[theorem]{Corollary}
\newtheorem{rem}[theorem]{Remark}
\newenvironment{prf}{\trivlist \item [\hskip
\labelsep {\bf Proof:}]\ignorespaces}{\qed \endtrivlist}

\newfont{\tenbi}{cmbxti10}
\newcommand{\beq}{\begin{equation}}
\newcommand{\eeq}{\end{equation}}
\newcommand{\rR}{\mathrm{R}}
\newcommand{\rQ}{\mathrm{Q}}
\newcommand{\rP}{\mathrm{P}}
\newcommand{\rA}{\mathrm{A}}
\newcommand{\rB}{\mathrm{B}}
\newcommand{\rC}{\mathrm{C}}
\newcommand{\rx}{\mathrm{x}}
\newcommand{\rz}{\mathrm{z}}
\newcommand{\rd}{\mathrm{d}}
\newcommand{\Z}{{\mathbb Z}}
\newcommand{\C}{{\mathbb C}}

\newcommand\la{{\lambda}}
\newcommand\La{{\Lambda}}

\newcommand\al{{\alpha}}
\newcommand\be{{\beta}}
\newcommand\gam{{\gamma}}

\newcommand\si{{\sigma}}
\newcommand\lax{{\bf L}}

\newcommand\bv{{\bf v}}
\newcommand\bx{{\bf x}}

\DeclareMathOperator{\Prym}{Prym}
\DeclareMathOperator{\Jac}{Jac}

\DeclareMathOperator{\Nm}{Nm}

\author{Yuri N. Fedorov\thanks{Department of Mathematics I, Politechnic university of Catalonia, Barcelona,
Spain. ~~E-mail: Yuri.Fedorov@upc.edu} $\,$
and Andrew N.~W. Hone\thanks{School of Mathematics,
Statistics and Actuarial Science, University of
Kent, Canterbury CT2 7NF, U.K. ~~E-mail: A.N.W.Hone@kent.ac.uk}}
\title{Sigma-function solution to the general Somos-6 recurrence via hyperelliptic Prym varieties} 

\begin{document}
\maketitle

\begin{abstract}We construct the explicit solution of the initial value problem for sequences generated by the 
general Somos-6 recurrence relation, in terms of the Kleinian sigma-function of genus two.
For each sequence there is an associated genus two curve $X$, such that  iteration of the 
recurrence corresponds to translation by a fixed vector in the Jacobian of $X$. The construction is based 
on a Lax pair with a spectral curve $S$ of genus four admitting an involution $\upsigma$ with two fixed points, 
and the Jacobian of $X$ arises as the Prym variety Prym$(S,\upsigma)$. 
\end{abstract}

\section{Introduction}
Somos sequences are integer sequences generated by quadratic  recurrence relations, 
which can be regarded as nonlinear analogues of the Fibonacci numbers.
They are also known as Gale-Robinson sequences, and   as well as
arising from reductions of bilinear partial difference equations in the theory
of discrete integrable systems, they 
appear in
number theory, statistical mechanics, string theory and algebraic combinatorics \cite{bmpw,bps,recs,fzlaurent}. 

This article is concerned with the general form of the sixth-order recurrence
\begin{equation}\label{S-6}
 \uptau_{n+6}\uptau_n = \alpha \uptau_{n+5}\uptau_{n+1} + \beta \uptau_{n+4}\uptau_{n+2} + \gamma \uptau_{n+3}^2, 
\end{equation}
with three arbitrary coefficients $\alpha,\beta,\gamma$.
It was 
an empirical observation
of Somos \cite{somos} that in the case $\alpha=\beta=\gamma=1$ the initial values $\uptau_0 = \cdots = \uptau_5=1$
generate a sequence of integers (A006722 in Sloane's Online Encyclopedia of Integer Sequences), which begins 
\beq\label{classical} 
1,1,1,1,1,1,3,5,9,23,75,421,1103,5047,41783,281527,\ldots .
\eeq 
Consequently, the relation \eqref{S-6} with generic coefficients is referred to as the Somos-6
recurrence, and the corresponding sequence $(\uptau_n )$ as a Somos-6 sequence. 

The first proof that the original 
Somos-6 sequence (\ref{classical}) consists entirely of integers was an unpublished result of Hickerson (see \cite{gale}); 
it relied on showing that the Somos-6 recurrence has the Laurent property, meaning that the iterates are 
Laurent polynomials in the initial data with integer coefficients. To be precise, in the general case  the 
iterates satisfy 
$$ 
\uptau_n\in \Z [\uptau_0^{\pm 1}, \ldots ,\uptau_5^{\pm 1},\alpha,\beta,\gamma ]\qquad \forall n\in\Z, 
$$
which was proved by Fomin and Zelevinsky as an offshoot of their development of cluster algebras \cite{fzlaurent}. 
The latter proof made essential  use of the fact that (\ref{S-6}) is a reduction of the cube recurrence, 
a partial difference equation which is better known in the theory of integrable systems as  Miwa's equation, or the bilinear form of the discrete BKP equation (see \cite{bkp}, for instance).   In the general case $\al\be\gam\neq 0$, (\ref{S-6}) does not arise from mutations in a cluster algebra,
although it does appear in the broader framework of Laurent phenomenon algebras \cite{LP}.

As was found independently by several people (see e.g. \cite{H05,honetams,poorten,ps} and references),
the sequences generated by general bilinear recurrences of order 4 or 5  are associated with 
sequences of points on elliptic curves, and 
can be written in terms of the corresponding Weierstrass sigma-function.
It was shown in \cite{H10} that sequences $(\uptau_n )$ produced by (\ref{S-6}) are the first ones which 
go beyond genus one: in general, they are parametrized
by a sigma-function in two variables.
To be precise, given a genus 2 algebraic curve $X$ defined by the affine model
\begin{equation} \label{can_X}
 z^2 = \sum_{j=0}^5 \bar{c}_j \, s^j \qquad \mathrm{with} \quad \bar{c}_5=4
\end{equation}
in the $(s,z)$ plane,
 let $\sigma({\bf u})$ denote the associated Kleinian sigma-function with ${\bf u}=(u_1,u_2)\in {\mathbb C}^2$, as
described in \cite{Baker2} (see also \cite{BEL, BEL2}). It gives rise to
the Kleinian hyperelliptic functions $\wp_{jk}({\bf u} )= -\partial_j\partial_k \log\sigma({\bf u})$, which are meromorphic on the Jacobian variety $\Jac(X)$ and generalize the Weierstrass elliptic $\wp$ function.

\begin{theorem}[\cite{H10}]\label{sigform} 
For arbitrary $\rA,\rB,\rC\in {\mathbb C}^*,\, {\bf v}_0\in {\mathbb C}^2$, the sequence with $n$th term
\begin{equation} \label{sigma_sol}
 \uptau_n =\rA \rB^n \rC^{n^2-1}\frac{\sigma({\bf v}_0+n {\bf v} ) }{\sigma( {\bf v} )^{n^2} } 
\end{equation}
satisfies the recurrence \eqref{S-6} with coefficients
\begin{gather}  \alpha = \frac{\sigma^2(3{\bf v}) \rC^{10}  }{\sigma^2(2{\bf v} ) \sigma^{10}({\bf v})} \,\hat\alpha , 
\qquad
\beta =\frac{\sigma^2(3 {\bf v} ) \rC^{16} }{\sigma^{18}({\bf v})} \,\hat \beta, \notag \\
\gamma = \frac{\sigma^2(3{\bf v} ) \rC^{18} }{\sigma^{18}({\bf v})}
\left( \wp_{11} (3 {\bf v} ) -\hat\alpha \wp_{11} (2 {\bf v} ) - \hat\beta  \wp_{11} ( {\bf v} ) \right ) , \label{abg}
\end{gather}
where 
\beq 
\hat\alpha = \frac { \wp_{22} (3{\bf v} ) - \wp_{22} ({\bf v}) }{\wp_{22} (2{\bf v} ) - \wp_{22} ({\bf v}) } , 
\quad 
\hat\be = \frac { \wp_{22} (2{\bf v} ) - \wp_{22} (3{\bf v}) }{\wp_{22} (2{\bf v} ) - \wp_{22} ({\bf v}) }
=1- \hat\al, 
\label{abhat}
\eeq 
provided that ${\bf v} \in\Jac(X)$ satisfies the constraint
\begin{equation} \label{det3}
\det \begin{pmatrix} 1 & 1 & 1 \\
   \wp_{12} ( {\bf v} ) & \wp_{12}(2 {\bf v} ) &  \wp_{12}(3 {\bf v} ) \\
 \wp_{22}({\bf v} ) & \wp_{22}(2{\bf v} ) &  \wp_{22}(3{\bf v} ) \end{pmatrix} =0 . \quad
\end{equation}
\end{theorem}
The preceding statement differs slightly from that of Theorem 1.1 in  \cite{H10}, in that we have used an alternative 
(but equivalent) expression for $\hat\al$ in (\ref{abhat}), and have included an additional parameter $\rC$ which is needed in what follows. 
Now while the above result means that the expression (\ref{sigma_sol}) is a solution of  \eqref{S-6} with suitable coefficients, 
it does not guarantee that it is the general solution, in the sense that the sequence $(\uptau_n)$ can always be written in this way, for a generic choice of initial data and coefficients. 
The ultimate purpose of this paper is to show that this is indeed the case. Our main result is the solution of the 
initial value problem by explicit reconstruction of the parameters appearing in (\ref{sigma_sol}), which yields the following. 
\begin{theorem}\label{main} 
For a sequence of complex numbers $(\uptau_n)$ generated by the recurrence \eqref{S-6} with generic values 
of the initial data  $\uptau_0,\ldots,\ldots\uptau_5$ and coefficients $\al,\be,\gam$, there exists a genus 2 curve $X$ with 
affine model (\ref{can_X}) and period 
lattice $\La$, points $\bv_0,\bv\in  \Jac(X)\simeq \C^2\bmod\La$ with $\bv$ satisfying (\ref{det3}), and 
constants $\rA,\rB,\rC\in\C^*$ such that the terms and coefficients are parametrized by the corresponding Kleinian 
functions according to (\ref{sigma_sol}) and (\ref{abg}), respectively. 
\end{theorem}
In order to 
solve the reconstruction problem, 
it will be convenient to work with a reduced version of the Somos-6 recurrence.  
The parameters $\rA,\rB$ in   (\ref{sigma_sol}) correspond to the group of scaling symmetries 
$\uptau_n\rightarrow \rA\rB^n\uptau_n$, which maps solutions to solutions, and considering invariance 
under this symmetry leads to certain quantities $\rx_n$, as described in the next paragraph. 
The parameter $\rC$ corresponds to covariance under the further 
scaling $\uptau_n\rightarrow \rC^{n^2}\uptau_n$, which maps solutions of \eqref{S-6} to solutions of the same 
recurrence with rescaled coefficients; in due course we will consider  quantities 
that are also invariant with respect to this additional symmetry. 

\paragraph{The reduced Somos-6 map.} 
Sequences generated by iteration of the Somos-6 recurrence \eqref{S-6} are equivalent to the orbits of the birational
map
$$
\varphi\, :\, (\uptau_0,\uptau_1,\dots, \uptau_5 ) \mapsto (\uptau_1,\uptau_2,\dots, \uptau_6), \quad
\uptau_6 = \frac{1}{\tau_0}\left(\alpha \uptau_{5}\uptau_{1} + \beta \uptau_{4}\uptau_{2}+\gamma \uptau_{3}^2\right)\,.
$$
As was observed in \cite{H10}, this map is Poisson with respect to the 
log-canonical bracket $\{ \uptau_m,\uptau_n\}=(m-n)\uptau_m\uptau_n$, which has four independent Casimir functions
\beq\label{casi}
 \rx_j = \frac{\uptau_j\, \uptau_{j+2}}{\uptau_{j+1}^2}, \qquad j=0,\dots, 3;
\eeq
these quantities are also invariant under the scaling transformation $\uptau_n\rightarrow \rA\rB^n\uptau_n$. 
The map $\varphi$ induces a recurrence of order 4 for a corresponding sequence $(\rx_n)$, that is 
\begin{equation} \label{rec-4}
 \rx_{n+4}\, \rx^2_{n+3}\, \rx^3_{n+2}\, \rx^2_{n+1}\,\rx_n= \alpha \,\rx_{n+3}\,\rx_{n+2}^2\, \rx_{n+1} + \beta\, \rx_{n+2} + \gamma, 
\end{equation}
which is equivalent  to iteration of a
birational map $\hat\varphi$ in ${\mathbb C}^4$ with coordinates ${\mathbf x}=(\rx_0,\dots, \rx_3)$.
We will refer to $\hat\varphi$  as the reduced Somos 6 map.  

The map
$\hat\varphi$ defined by (\ref{rec-4}) preserves the meromorphic volume form
$$
   \hat V = \frac{1}{\rx_0 \rx_1\rx_2\rx_3 }\, \rd\rx_0 \wedge \rd\rx_2 \wedge\rd\rx_2 \wedge \rd\rx_3
$$
 for arbitrary values of $\alpha, \beta, \gamma$,
and has two independent rational first integrals, here denoted 
$ K_{1}({\mathbf x}), K_{2}({\mathbf x})$, which are presented 
explicitly in section 2 below.
According to \cite{H10}, the map $\hat\varphi$ is also integrable in the Liouville--Arnold sense \cite{maeda}, 
at least in the case
$\alpha\beta\gamma =0$. In this paper are concerned with the general case $\alpha\beta\gamma \neq 0$, where 
a symplectic structure for the map $\hat\varphi$ is not known. 

On the other hand, a genus 2 curve \eqref{can_X} and the corresponding
sigma-function solution \eqref{sigma_sol}, \eqref{abg} of the Somos-6 map $\varphi$ 
imply that 
the   solution of \eqref{rec-4} is 
\begin{equation} \label{x_n}
 \rx_n = \frac{ \rC^2\sigma( {\bf v}_0 + n{\bf v} )\, \sigma({\bf v}_0 +(n+2) {\bf v} ) }
{\sigma^2({\bf v}_0 +(n+1) {\bf v} )\, \sigma^2({\bf v}) }.
\end{equation}
In view of the addition formula for the genus 2
sigma-function \cite{Baker2}, the right hand side of \eqref{x_n} can be written in terms of Kleinian $\wp$ functions as 
\begin{equation} \label{x_wp}
\rx_n = \rC^2\Big(\wp_{22} ({\bf u}) \wp_{12}({\bf v}) - \wp_{12} ({\bf u}) \wp_{22}({\bf v}) + \wp_{11}({\bf v}) -\wp_{11}({\bf u})\Big)=:{\cal F}({\bf u}),
\end{equation}
where ${\bf u} ={\bf v}_0 + (n+1){\bf v}$. Note that, when ${\cal F}$ is considered as a function on the Jacobian, 
${\cal F}({\bf u})$ is singular if and only if 
${\bf u}\in (\sigma)$, the theta divisor in $\Jac(X)$ (using the notation in \cite{BEL}). 
Then, upon setting $n=0$, we have the map 
\beq\label{uppmap}
\uppsi: \quad {\bf u} \mapsto 
\Big({\cal F}({\bf u}),{\cal F}({\bf u+v}),\mathcal{F}({\bf u}+2{\bf v}),{\cal F}({\bf u}+3{\bf v})\Big) 
= (\rx_0,\rx_1,\rx_2,\rx_3), 
\eeq
which is 
a meromorphic embedding $\uppsi:\,\Jac(X)\setminus (\sigma_{0123})\to \mathbb{C}^4$, where 
$(\sigma_{0123})$ denotes the theta divisor $(\sigma)$ together with its translates by ${\bf v},2{\bf v}$ and $3{\bf v}$. 
%
Once Theorem \ref{main} is proved (see section 6), 
we are able to recover 
$\rC$, $\bv$ and ${\bf u}=\bv_0 +\bv\in\Jac(X)$ from the coefficients and initial data of the map,  
so that we arrive at

\begin{theorem} \label{Kum}
Generic complex invariant manifolds ${\cal I}_K =\{ K_1({\mathbf x})=k_1,  K_2({\mathbf x})=k_2\} $ of the map 
$\hat\varphi$ are isomorphic
to open subsets of $\Jac(X)$.
\end{theorem}

For the purposes of our discussion, it will be more convenient to describe the reduced Somos-6 map in an alternative set of coordinates. We introduce the quantities 
\beq 
P_n=-\frac{\delta_1\be}{\rx_n\rx_{n+1}}, \qquad 
R_n=\frac{\delta_1\gam}{\rx_n\rx_{n+1}\rx_{n+2}},\qquad \mathrm{with} \quad 
 \delta_1 = \sqrt{ - \frac{\alpha}{\beta \gamma}}, 
\label{coord_PR}
\eeq
so that $\rx_n=-\gam P_{n-2}/(\be R_{n-2})$, and $P_0,P_1,R_0,R_1$  
are birationally related to $\rx_0, \rx_1, \rx_2, \rx_3$.  Thus, after  conjugating 
$\hat\varphi \, : {\mathbb C}^4 \to {\mathbb C}^4$ by a birational change of variables,  
we can rewrite it in the form 
$(P_0,P_{1}, R_0,R_{1})\mapsto (\tilde{P}_0,\tilde{P}_{1}, \tilde{R}_0,\tilde{R}_{1})$,
where  
\beq\label{rpmap} 
\tilde{P}_0 = P_1, \quad \tilde{P}_1 = \frac{\mu R_0 R_1}{P_0P_1}, \quad 
\tilde{R}_0 = R_1,\quad 
\tilde{R}_1 = (P_0+\la R_0R_1-\la P_0R_1)^{-1},
 \eeq 
with the coefficients 
\beq\label{lamu}  
\la = \frac{\delta_1\beta^2}{\alpha^2}, \qquad 
\mu = -\frac{\delta_1\beta^3}{\gamma^2}. 
\eeq
Observe that, from the analytic formulae (\ref{abg}) and (\ref{x_n}), the quantities $P_n,R_n$ and the coefficients 
$\la,\mu$ are independent of the parameter $\rC$.


\paragraph{Outline of the paper.}
In the next section, we describe the first of our main tools, namely the $3\times 3$ Lax pair 
for the map $\hat\varphi$, which (as announced in \cite{H10}) is
obtained from the associated Lax representation for the discrete BKP equation. The corresponding spectral curve $S$ yields 
the first integrals $ K_{1}, K_{2}$.  
However, $S$ is not the required genus 2 curve $X$, but rather it is trigonal
of genus 4, having an involution $\upsigma$ with two fixed points. Then it turns out that the 2-dimensional Jacobian of $X$,
which is the complex invariant manifold of the map $\hat\varphi$ according to Theorem \ref{Kum},
can be identified with the Prym subvariety $\Prym(S,\upsigma)$ of $\Jac(S)$. (An analogous situation was described 
recently for an integrable H\'{e}non-Heiles system \cite{EFH15}.) 

To obtain an explicit algebraic description of $\Prym(S,\upsigma)$ and, therefore, of the curve $X$, we make use 
of recent work
by Levin \cite{Levin} 
on the general case of double covers of hyperelliptic curves with two branch points. 
All relevant 
details are given 
in section 3.

In section 4 it is shown how the discrete Lax pair allows a description of the map $\hat\varphi$ as a translation on
$\Prym(S,\upsigma)\subset\Jac(S)$ by a certain vector. 
This translation 
is subsequently identified with a specific
degree zero divisor on $X$ representing the required vector $\bv\in\Jac(X)$, and in section 5 
we also explicitly find degree zero divisors on $X$ representing the vectors $2{\bf v}, 3{\bf v}$.
This enables us to rewrite the determinantal constraint \eqref{det3} in terms of the above three divisors, and then
observe that it is trivially satisfied.

In section 6, all of the required ingredients are ready to present the reconstruction of  the sigma-function solution (\ref{sigma_sol}) from the initial data and coefficients, which proves Theorem \ref{main}. We also provide a couple of 
explicit examples, including the original Somos-6 sequence (\ref{classical}).  The paper ends with some conclusions, followed by 
an appendix which includes the derivation of the Lax pair and another technical result. 

\section{The Lax pair, its spectral curve, and related Jacobian varieties}
\setcounter{equation}{0} 

The key to the solution of the initial value problem for the Somos-6 recurrence is the Lax representation 
of the map $\hat\varphi$. 

\begin{theorem} \label{laxp} 
The mapping $\hat\varphi \, : {\mathbb C}^4 \to {\mathbb C}^4$
is equivalent to the discrete Lax equation 
\beq\label{dlax} 
\tilde{{\bf L}}{\bf M}={\bf M}{\bf L}, 
\eeq 
with 
\beq\label{laxm}
{\bf L}(x) = \left(\bear{ccc} 
\dfrac{A_2x^2+A_1x}{x+\la} & \dfrac{A_2'x^2+A_1'x}{x+\la} &\dfrac{A_1''x+A_0''}{x+\la} \\ 
B_2\, x^2+B_1 x & B_1' x & B_1'' \, x+B_0'' \\ 
C_2x^2+C_1x & C_2'x^2 + C_1' x & C_1''x+C_0''
\eear\right),
\eeq
\beq\label{Lax3} 
{\bf M}(x) =\frac{1}{R_0}
\left(\bear{ccc} -1 & 1 & 0 \\ 
-\frac{x}{\la}-1 &1 & \frac{1}{\la} \\ 
0 & (\la P_0R_{1}R_{2}+1)x & -P_0R_{2} 
\eear\right), 
\eeq 
where $R_2=\tilde{R}_1$ as in (\ref{rpmap}), and 
$$ \bear{rl} 
A_1 =& P_0\left(P_1+\dfrac{1}{R_1}-\dfrac{P_1}{R_0R_1} \right) 
+\mu\left( \dfrac{(R_0P_1+1)R_1}{P_0P_1} -R_0-R_1\right) \\ 
& +\la\left(R_0+R_1-P_0-P_1 +P_1R_0R_1-P_0P_1R_1 +\dfrac{P_0P_1-1}{R_0}
\right) 
\\
& 
+\la\mu\left(R_0R_1 -\dfrac{(P_1R_0+1)R_0R_1}{P_0P_1}\right) 
+\dfrac{\mu}{\lambda} , 
\eear 
$$
$$  \bear{rl} 
A_2 =&  P_0+P_1+\la (R_0-P_0)R_1 +\dfrac{1}{R_0 R_1}\left( R_0-P_1-\dfrac{1}{\la}\right) 
\\ &
+\mu\dfrac{ (R_1-P_0-P_1)R_0-P_0R_1}{P_0 P_1}
+\dfrac{\mu}{\la}\left(\dfrac{1}{P_0}+\dfrac{1}{P_1}\right) 
+\la\mu\dfrac{(P_0-R_0)R_0 R_1}{P_0P_1}, 
\eear $$ 
$$ \bear{rl} 
A_1' = &\la \left( P_0-R_1-P_0P_1R_1+\dfrac{1-P_0P_1}{R_0}\right) 
+\mu\left(1-\dfrac{1}{P_0P_1}\right) R_1
\\ &
+\la\mu\left(\dfrac{1}{P_0P_1}-1 \right) R_0R_1,\qquad 
A_2' = \la P_0R_1 +\mu\dfrac{R_1}{P_1} -\la\mu \dfrac{R_0R_1}{P_1},
\eear
$$ 
$$ 
A_0''= \frac{P_0P_1}{R_0R_1}-P_0P_1-\frac{P_0}{R_1}+\mu R_0 
-\frac{\mu}{\lambda}, 
$$ 
$$ 
A_1'' = \frac{P_1}{R_0R_1}-\frac{1}{R_1}-P_0-P_1 
+\mu \left(R_0-\frac{1}{\la}\right)\left(\frac{1}{P_0}+\frac{1}{P_1}\right) 
+\frac{1}{\la R_0R_1} 
,$$ 
$$ B_1 = 
R_0-P_0-P_1+\frac{P_0P_1-1}{R_0}
+\frac{1}{\la}\left(\frac{P_0(R_0-P_1)}{R_0R_1}\right), 
\qquad
B_2=-\frac{P_1}{\lambda R_0R_1},  
$$
$$ 
B_1'= P_0+\frac{1-P_0P_1}{R_0}, \qquad
B_0'' = \frac{P_0(P_1-R_0)}{\la R_0R_1}, \qquad 
B_1'' = \frac{P_1}{\la R_0R_1}, 
$$
$$ C_1= 
\mu\left(\frac{R_0P_1+1}{P_0P_1}-1\right)R_1 +\frac{\mu}{\lambda}
,$$ 
$$ 
C_2 = P_0 +\la (R_0-P_0)R_1 +\mu\frac{(R_0-P_0)R_1}{P_0P_1} - \frac{1}{\la R_0R_1}  
+ \frac{\mu}{\lambda}\left(\frac{1}{P_0}+\frac{1}{P_1}\right) 
,$$ 
$$ 
C_1'= \mu\left(1-\frac{1}{P_0P_1} \right) R_1
, \qquad 
C_2' = \la P_0R_1+\mu\frac{R_1}{P_1},
$$ 
$$ 
C_0'' = -\frac{\mu}{\lambda}
,\qquad 
C_1''= -P_0+\frac{1}{\la R_0R_1}  - \frac{\mu}{\lambda}\left(\frac{1}{P_0}+\frac{1}{P_1}\right) 
.
$$  
\end{theorem}
\begin{prf} The equation (\ref{dlax}) can be checked directly with computer algebra. 
For the rather more  straightforward origin of this complicated-looking Lax pair, 
see the appendix.   
\end{prf}

The characteristic equation  $\det ({\bf L}(x)-y\mathbf{1})$ 
defines the spectral curve $S \subset {\mathbb C}^2(x,y)$, which,  
after elimination of the common factor $1/(x+\la)$, is given by 
\begin{equation} \label{spectral}
 f(x,y):= (x + \lambda )\,y^{3} + (x\,{K_{1}} + \mu  +
x^{2}\,{K_{2}})\,y^{2} - (\mu \,x^{4} + {K_{1}}\,x^{3} + x^{2}\,{K_{2}})\,y - \lambda \,x^{4} - x^{3} =0 ,
\end{equation}
where 
$K_1, K_2$ are independent first integrals, namely
\begin{align}
  & \quad K_1 =\frac{\hat{K}_1 (P_0,P_1, R_0, R_1) }{P_0 P_1 R_0 R_1 }, \quad
K_2 =\frac{\hat{K}_2 (P_0,P_1, R_0, R_1) }{P_0 P_1 R_0 R_1 }, \label{ints_Kj} \\
\text{with}\qquad\qquad\, 
\hat{K}_1 & = \la \mu R_{0}^2 R_{1}^{2} ({R_{0}}\,{P_{1}} - {P_{0}}\,{P_{1}} + 1) \notag \\
 & \quad + \la {R_{0}}\,{R_{1}}\,{P_{0}}\,{P_{1}}\,
({R_{1}}\,{P_{0}}\,{P_{1}} + {P_{0}} - P_{1} R_{0} R_{1} - {R_{1}}+ {P_{1}}- R_{0}) \notag \\
& \quad - \mu R_{0} R_{1} ({R_{1}} - {P_{0}} - {R_{1}}\,{P_{0}}
\,{P_{1}} - {R_{0}}\,{P_{0}}\,{P_{1}} - {P_{1}} + {P_{1}}\,{R_{0} }\,{R_{1}})  \notag \\
& \quad - {P_{0}}\,{P_{1}}\,({P_{0}}\,{P_{1}}\,{R_{0}}\,{R_{1}}
 - {P_{0}}\,{P_{1}} + 1 + {P_{0}}\,{R_{0}})  ,\notag \\
\hat{K}_2 & = \la \,\mu  {R_{0}}^{2}\,{R_{1}}^{2}\,(R_{0} - P_{0})
- \la R_{0} R_{1}^{2} {P_{0}} {P_{1}}\,(R_{0} - P_{0}) \notag \\
 & \quad + \mu \, R_{0}\,R_{1}\,(P_{0} R_{1} - R_{1}\,R_{0} + P_{0} R_{0} +R_{0}\,P_{1}) \notag \\
 & \quad - {P_{0}}\,{P_{1}}\,( - {R_{1}}\,{P_{0}}\,{P_{1}} + P_{0}\,{R_{0}}\,{R_{1}} +
P_{1}R_{0} R_{1} + R_{1} - P_{1} + R_{0}). 
\end{align}
\begin{rem} 
Replacing the variables $P_0,P_1, R_0, R_1$ by the expressions \eqref{coord_PR} yields the first integrals
 $ K_1(\bx),  K_2(\bx)$ of the reduced map $\hat \varphi$ in the original variables $\rx_0,\dots, \rx_3$. 
These are seen to be 
rescaled versions of the quantities $H_1,H_2$ derived in \cite{H10} from higher order 
bilinear relations, according to 
\beq\label{h1h2}
K_1=\frac{\be H_1}{\al\gamma^2}, \qquad 
K_2=\frac{\delta_1\be H_2}{\al\gamma}. 
\eeq 

One can also verify that, for generic values of $\la, \mu, k_1, k_2$, the complex invariant manifold ${\cal I}_K =\{ K_1({\mathbf x})=k_1,  K_2({\mathbf x})=k_2\}$ is irreducible.  
\end{rem} 
\medskip  

The curve $S$ is trigonal of genus 4 and has an interesting
involution $\upsigma\, :\; (x,y)\to (1/x, 1/y)$ with two fixed points, namely $(1,1)$ and $ (-1,1)$.

We compactify $S$ by embedding it in ${\mathbb P}^2$
with homogeneous coordinates  $(X:Y:Z)$, where $x=X/Y, y=Y/Z$. The compact curve has a singularity at
$(0:1:0)$. After regularization, this point gives two points at infinity: the first one is 
$ \bar{\cal O}=(x=\infty,y=\infty)$ with the Laurent expansion
$$
x= \frac{1}{\tau^2}+O(\tau^{-1}), \quad y= \frac{1}{\sqrt{\mu}\, \tau^3} +O(\tau^{-2})$$
with respect to a local parameter $\tau$ near $\bar{\cal O}$;
and the second is $\bar{\cal O}_2 =(x=-\la, y=\infty)$, with the Laurent expansion
$$
x= -\la +O(\tau), \quad y= -\frac{\la}{\tau} +O(1) .
$$
The third point  at infinity $\bar{\cal O}_1= (x=\infty, y=-\la/\mu)$ comes from $(1:0:0)$ and has the expansion
$$
x= \frac{1}{\tau}+O(1), \quad y= -\frac{\la}{\mu} + O(\tau) .
$$
Under the action of  $\upsigma$, these points are in involution with the following three finite points:
\begin{align*}
{\cal O} & =(x=0,y=0) \quad \text{with} \quad x= \tau^2, \quad y=\frac{\tau^3}{ \sqrt{\mu}} +O(\tau^4) ; \\
{\cal O}_1 &= (x=0,y=-\mu/\la) \quad \text{with} \quad x= \tau; \\
{\cal O}_2 & = (x=-1/\la,y=0) \quad \text{with} \quad y=\tau +O(\tau^2) .
\end{align*}

The above three pairs of points on $S$ will play an important role, so we depict them on the 
diagram above, 
with arrows denoting the involution $\upsigma$.
\begin{figure}[h,t]
\begin{center}
\includegraphics[width=0.75\textwidth]{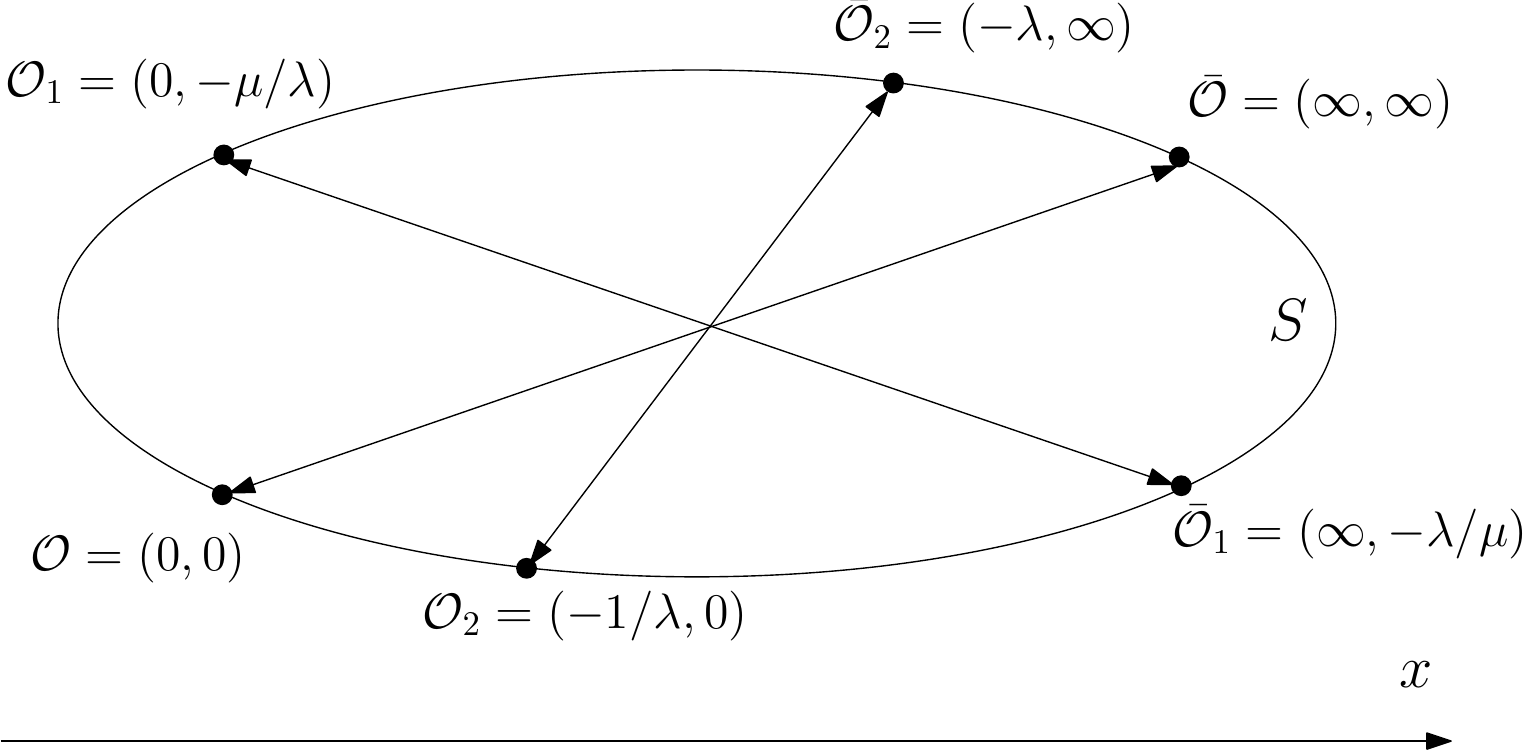}
\end{center} 
\end{figure}
The curve $S$ can be viewed as a 3-fold cover of ${\mathbb P}^1$ with affine coordinate $x$.
As follows from the above description, the points ${\cal O}, \bar{\cal O}$
are ordinary branch points of the covering, and
there is no branching at the points ${\cal O}_1, \bar{\cal O}_1, {\cal O}_2, \bar{\cal O}_2$.
It follows that the divisors of zeros and poles of the coordinates $x,y$ on $S$ are
\begin{equation} \label{div_x_y}
   (x) = 2{\cal O} + {\cal O}_1 - 2\bar{\cal O} - \bar{\cal O}_1, \quad
(y) = 3{\cal O} + {\cal O}_2 - 3\bar {\cal O} - \bar {\cal O}_2 .
\end{equation}

 Observe that a generic complex 2-dimensional invariant manifold ${\cal I}_K$ of the reduced Somos-6 map
$\hat\varphi$ cannot be the Jacobian of $S$, as the latter has genus 4.
The curve $S$ is a 2-fold covering of a curve $G=S/\upsigma$ whose genus is 2, by the Riemann--Hurwitz formula.
The involution $\upsigma$ extends to $\Jac(S)$ which then contains two Abelian
subvarieties: the Jacobian of $G$, which is invariant under $\upsigma$, and the 2-dimensional Prym variety, denoted
$\Prym(S, \upsigma)$, which is anti-invariant with respect to $\upsigma$. It will play a key role in the
description of the complex invariant manifolds of the map $\hat\varphi$.  For this purpose it is
convenient to recall some properties of Prym varieties corresponding to our case.

\section{Hyperelliptic Prym varieties}
\setcounter{equation}{0} 

\paragraph{Generic double cover of a hyperelliptic curve with two branch points.}
Consider a  genus $g$ hyperelliptic curve $C$: $v^2=f(u)$, where
$f(u)$ is a polynomial of degree $2g+1$ with simple roots.
As was shown in \cite{Levin}, any double cover of $C$ ramified at two finite points
$P=(u_P,y_P), Q=(u_Q, y_Q) \in C$ (which are not related to each other by the hyperelliptic involution on $C$, i.e.,
$u_P \ne u_Q$)
can be written as a space curve of the form 
\begin{equation} \label{big_CC}
{\widetilde C} \, : \; \rz^2 = v + h(u), \quad v^2=f(u),
\end{equation}
where $h(u)$ is a polynomial of degree $g+1$ such that
$$
h^2(u)-f(u) = (u-u_P) (u-u_Q) \rho^2(u)
$$
with $\rho(u)$ being a polynomial of degree $g$. (Here $u_P$ or $u_Q$ may or may not coincide with roots of $\rho(u)$.)
Thus $\widetilde C$ admits the involution $\upsigma\, : \; (u,v,\rz) \mapsto (u,v,-\rz)$, with fixed points
$(u_P,y_P,0), (u_Q, y_Q,0) \in {\widetilde C} $.
Then the genus of $\widetilde C$ is $2g$, and it was shown by Mumford  \cite{Mum} and Dalaljan \cite{Dal} 
that 
\begin{itemize}
\item
$\Jac( \widetilde C )$ contains two $g$-dimensional Abelian subvarieties: Jac$(C)$ and the
Prym subvariety $\Prym ({\widetilde C},\upsigma)$, with the former 
 invariant under the extension of  $\upsigma$ to $\Jac(\widetilde C)$, and the latter
anti-invariant; 
\item
Prym$({\widetilde C},\upsigma)$ is principally polarized and 
is the Jacobian of a {\it hyperelliptic} curve $C'$.
\end{itemize}
It was further shown recently by 
Levin \cite{Levin} that the second curve $C'$ can be written explicitly as
\begin{equation} \label{classical_C'}
w^2 = h(u)+ Z, \quad Z^2=h^2(u)-f (u)\equiv (u-u_P) (u-u_Q) \rho^2(u) ,
\end{equation}
which is equivalent to the plane curve $w^4-2 h(u)w^2 + f(u) =0$.
The latter can be transformed explicitly to a hyperelliptic form by an algorithm given in \cite{Levin}.

In order to apply the above results to obtain an explicit description of the Prym variety 
$\Prym(S, \upsigma)\subset \Jac(S)$ in
our case, we will need
\begin{proposition} 1) The quotient of $S$ by the involution $\upsigma$ is the genus 2 curve $G\subset {\mathbb C}^2(T,Y)$
given by the  equation
 \begin{equation} \label{gen_2}
G (T,Y):=  A Y^3 + B(T) Y^2 + [C(T)-3A] Y + D(T) =0 ,
\end{equation}
where
$A =-\mu^2$, $B(T) = \mu \,( - {K_{2}}\,T^{2} + T\,{K_{1}} + 2\,\lambda  + 2\,{K_{2}} - T^{3} + 3\,T)$, 
\begin{align*}
C(T) & = \lambda \,T^{4}\,\mu  + ({K_{1}}\,\mu  + \lambda )\,T^{3} +
(\lambda \,{K_{2}} - 4\,\mu \,\lambda  - {K_{2}}\,\mu  + {K_{1}})\,T^{2} \\
& \mbox{} + ({K_{1}}\,{K_{2}} - 3\,\lambda  - {K_{2}} - \lambda \,{
K_{1}} - \mu  - 3\,{K_{1}}\,\mu )\,T + 2\,\mu \,\lambda  - \lambda ^{2} - ({K_{1}}^{2}+  {K_{2}}^{2}+1)
   \\
& \mbox{}  - 2({K_{1}}+ \lambda \,{K_{2}} - \mu\,{K_{2}}) , \\
D(T) & = (\mu ^{2} + \lambda ^{2})\,T^{4} + 2\,\lambda \,T
^{3}\,{K_{1}} + ({K_{1}}^{2} - 4\,\lambda ^{2} - 4\,\mu ^{2} - 2
\,\lambda \,{K_{2}} + 1)\,T^{2} \\
& 
\mbox{} + ( - 2\,{K_{1}}\,{K_{2}} - 6\,\lambda \,{K_{1}} + 2\,{K_{2}}
- 2\,\lambda ) T + 2(\lambda ^{2}+\mu ^{2}) \\ 
& - 2({K_{1}}^{2}  -{K_{2}}^{2}+1)
 - 4({K_{1}}  - \lambda {K_{2}}).
\end{align*}
The double cover $\pi \, : \, S\to G$ is described by the relations
\begin{equation} \label{TY}
 T= \frac{x}{y}+ \frac{y}{x} , \quad Y=y+ \frac 1y \, ,
\end{equation}
and the images of the branch points $P=(x=1,y=1), Q=(x=-1,y=1) \in S$ on $G$ are
$P=(T=2,Y=2), Q=(T=-2,Y=2)$.
\medskip

\noindent
2)  The curve $G$ is equivalent to the following  curve $C \subset {\mathbb C}^2(u,v)$ in hyperelliptic form: 
\begin{align}
C \, : \quad v^{2} & = \rP_6(u), \label{Gam} \\
\rP_6(u) & =  1 + (4\,\mu  - 2\,{K_{2}})\,u
 + ( 4\,\mu \,\lambda+ {K_{2}}^{2}   - 2\,{K_{1}} )\,u^{2} \notag  \\
& \quad + (   2\,\lambda - 10\,\mu  + 2\,{K_{1}}\,{K_{2}} )\,u^{3
} + ( - 8\,\mu \,\lambda   + {K_{1}}^{2}  - 2\,\lambda \,{K_{2}}  + 2\,\mu\,{K_{2}})\,u^{4}  \notag \\
& \quad + (  4\,\mu- 2\,\lambda\,{K_{1}}  + 2\,\mu\,{K_{1}} )\,u^{5} + (\mu + \lambda)^2\,u^{6} \, . \notag
\end{align}
The birational transformation between $G$ and $C$ is described by the relations
\begin{align}
T  & =  - \frac {1}{2} \,
\frac {(\la+\mu)u^3 + (2+K_1)u^2+ (K_2+2\la)u+1-v}{u\,(1 + \lambda \,u)},  \label{TT} \\
Y  & =  - \frac {1}{2 \mu \,u^{3}(1 + \lambda \,u)} \bigg [ (\la-\mu)^2 u^4 - (\la-\mu)(K_1-1) u^3 \notag \\
& \quad + ((\mu-\la) K_2 + 4\la \mu -K_1)u^2 +(\la+3\mu-K_2) u +1 - (1+\mu u+\la u) v \bigg ] . \label{YY}
\end{align}
The branch points $P=(T=2,Y=2), Q=(T=-2,Y=2)\in G$ on $C$ are, respectively,
$P= (u_P,v_P) , Q=(u_Q,v_Q)$ with
\begin{gather}
  u_P = -\frac {F_2}{F_1} : =- \frac {2 + \lambda - \mu  + K_2}{2\,\lambda+ 1 + 2\,\mu+ K_1}
\quad
u_Q = \frac {\bar F_2}{\bar F_1} :=\frac {- 2 + \lambda - \mu  + K_2}{2\,\lambda-1+2\mu-K_1},
\label{uPQ} \\
\begin{aligned}
v_P & = \frac{1}{F_1^3} \left( F_1^3- (2+\la+\mu) F_1^2 F_2 + (2\la-2\mu+1) F_1 F_2^2-(\la+\mu) F_2^3 \right)  ,  \\
v_Q & = \frac{1}{\bar F_1^3}
\left(\bar F_1^3 + (-2+\la+\mu) \bar F_1^2 \bar F_2 + (-2\la+2\mu+1) \bar F_1 \bar F_2^2-(\la+\mu) \bar F_2^3 \right)\, .
\end{aligned}
\end{gather}
\end{proposition}
\begin{prf} Applying the substitution \eqref{TY} to the polynomial $G(T,Y)$ we get the  product 
 $f(x,y) \tilde{f}(x,y)/(x^{4}\,y^{2})$, where 
\begin{gather*}
\tilde{f} = 
(x^{3}\,y^{2} - x\,y^{5})\,K_{1} + (x^2\,y^4 -x^2 y^3) K_2
+ \lambda x^4 \,y + \mu \,x^{4}- x\,y^{4} + y^{3}\,x^{3}- y^{7}\,\mu - \lambda \,y^{6} , 
\end{gather*}
and this product  is zero due to the equation \eqref{spectral}. Hence $T,Y$ satisfy $G(T,Y)=0$.
The proof of the other items is a direct calculation (which we made with Maple). 
\end{prf} 

$\Prym(S,\upsigma)$ is isomorphic to the Jacobian of a second genus 2 curve $C'$, and
in order to find its equation by applying
the algorithm of \cite{Levin} described above,
it is convenient to represent the curve $S$ in a form similar to \eqref{big_CC}.

\begin{proposition} \label{can_form}
The spectral curve $S$ is equivalent to the space curve 
\begin{equation} \label{tC}
\tilde C \,:\;  v^2 = \rP_6(u), \quad w^2 =4 \mu u^6 (1+\la u)^2\, (Y(u,v)-2)(Y(u,v)+2) \equiv h(u)+ g(u)v,
\end{equation}
where $\rP_6(u)$ is given by \eqref{Gam}, $Y(u,v)$ by \eqref{YY}, and
$g(u), h(u)$ are polynomials of degree 5 and 8 respectively, obtained by replacing $v^2$ in the right hand side
of the second equation by $\rP_6(u)$.
On $\tilde C$ the involution $\upsigma$ is given by $(u,v,w) \to (u,v,-w)$, and its fixed points are
$P=(u_P,v_P,0), Q=(u_Q,v_Q,0)$.
\end{proposition}

Explicit expressions for $h(u), g(u)$ are relatively long and are not shown here.
Observe that $h(u), g(u), \rP_6(u)$ do not have common roots and that
\begin{gather}
h^2(u)- g^2(u) \rP_6(u)
= 16\,\mu ^{2}\,u^{6}\,(1 + \lambda \,u)^{2} (\bar F_2-u \bar F_1)(F_2+ u F_1)
\cdot Q^2(u) \label{ZZ}, \\
Q(u)=2(\la-\mu)\,u^{2} + (1-K_1)u -K_2-\la +\mu , \notag
\end{gather}
which, in view of the expressions \eqref{uPQ} for $u_P, u_Q$, yields
\begin{gather} \label{disc}
h^2(u)- g^2(u) \rP_6(u) =  F_1 \bar F_1\,(u-u_P)(u-u_Q)\,u^{6}\,(1 + \lambda \,u)^{2}\,Q^2(u).
\end{gather}
Thus, the function $w^2$ in \eqref{tC} is meromorphic on the hyperelliptic
curve $C$ and has simple zeros only at $P,Q$, and even order zeros elsewhere. It also has only even order poles at
the two points at  infinity on $C$. Hence, $\tilde C$ is a double cover of $C$ ramified at $P,Q$ only, as expected.

The above description of the curves and coverings can be briefly summarized in the diagram below, where the horizontal equals signs  denote birational equivalence.
$$
\begin{CD}
S  @= \tilde C \\
@ V 2:1 VV @ VV 2:1 V \\
G  @ = C
\end{CD}
$$
\medskip

\noindent{\bf Proof of Proposition \ref{can_form}:}
In view of relations \eqref{TY}, \eqref{YY}, the curve $S$ 
can be written as
$$
   v^2 = \rP_6(u), \quad \frac 1y +y = Y(u,v) \; \Longleftrightarrow 2 y= Y(u,v) + \sqrt{ Y^2 (u,v)-4 }.
$$
Under the birational transformation $(u,v,y) \to (u,v,\hat w=2y-Y(u,v))$, the above reads
$$
v^2 = \rP_6(u), \quad \hat w^2 = Y^2 (u,v)-4 .
$$
Then by the  substitution $\hat w=w/(2\mu u^3 (1+\la u))$, the latter 
equations 
take the form \eqref{tC}.
Next, since $\upsigma$ leaves the curve $C$ invariant, it does not change the coordinates $u,v$, so it only flips
the sign of $w$. Finally, since $Y(P)=Y(Q)=2$, from \eqref{tC} we get $w(P)=w(Q)=0$. \qed 
\medskip

Observe that the equation \eqref{tC} of $\tilde C$ does not have the same structure as that of the model curve
\eqref{big_CC}: the degrees of the corresponding polynomials do not match. Hence, the
formula \eqref{classical_C'} for the second hyperelliptic curve $C'$ 
is not directly applicable
to \eqref{tC}. For this reason, below we adapt the derivation of \eqref{classical_C'} to our situation.

\paragraph{Tower of curves and Jacobians.} Following the approach of \cite{Dal}, consider the tower of curves
\begin{equation} \label{tower}
\begin{CD}
 C @ < \pi << \widetilde C @ < \tilde \pi << \widetilde{\widetilde C} @ > \pi_1 >> C'
\end{CD}
\end{equation}
where $\widetilde C$ is given by \eqref{tC} and $\widetilde{\widetilde C}$ is a double 
cover of $\tilde C$ given by
\begin{equation} \label{big_C}
    \tilde{\tilde C} \; : \; v^2 = \rP_6(u), \quad {w}^2 = h(u) + g(u) v, \quad {\bar w}^2 = h(u) - g(u) v .
\end{equation}
The covering $\widetilde{\widetilde C} \to {\widetilde C}$ is ramified at the points on ${\widetilde C}$ 
where the function ${\bar w}^2$ has {\it simple} zeros or poles.
As shown above, the function $w^2 = h(u) + g(u) v$ has precisely 2 simple zeros
$P= (u_P,v_P), Q=(u_Q, v_Q)$ and no simple poles on $C$. Hence,
$h(u) - g(u) v$ has only two simple zeros $\bar P= (u_P,-v_P), \bar Q=(u_Q, -v_Q)$ on $C$. Since
$\tilde C$ 
is a double cover 
of $C$ and $\bar P, \bar Q$ are not branch points of ${\widetilde C}\to C$,
the function ${\bar w}^2$ has four  simple zeros on ${\widetilde C}$, namely
$\bar P,\bar Q, \upsigma(\bar P), \upsigma(\bar Q)$.
Hence $\widetilde{\widetilde C} \to {\widetilde C}$ is ramified at the latter four points, and so,
by the Riemann--Hurwitz formula, the genus of $\widetilde{\widetilde C}$ equals $9$.

The ``big" curve $\widetilde{\widetilde C}$ has various involutions, one of which is
$$
\upsigma_1 \; : \; (u,v,w,\bar w)\to (u,- v,\bar w, w) ,
$$
and the last curve in the tower \eqref{tower} is the genus 2 quotient curve
$C'= \widetilde{\widetilde C}/\upsigma_1$. The corresponding
projection $\widetilde{\widetilde C}\to C'$ is denoted $\pi_1$.
The projections $\tilde \pi$ and $\pi_1$ are explicitly given by
\begin{equation} \label{divs}
  \tilde\pi (u,v,w,\bar w) = (u,v,w), \quad  \pi_1 (u,v,w,\bar w ) = (u, W= (w+ \bar w)/\sqrt{2}, Z= w \cdot \bar w ).
\end{equation}

The tower \eqref{tower} is a part of a tree of curves introduced in \cite{Dal} for the general case of a 
genus $g$ hyperelliptic curve $C$.
As was shown in \cite{Dal}, the tree of curves implies relations between the corresponding Jacobian varieties described
by the following diagram,
$$ \begin{CD}
  \Jac(C') @ > \pi_1^* >> \Jac( \widetilde{\widetilde C} ) \\
      @|          @ A \tilde\pi^* AA \\
   \Prym(\tilde C,\upsigma) @ >>> \Jac(\tilde C) \\
      @.        @ A \pi^* AA \\
     @.    \Jac(C)
\end{CD} $$
where arrows denote inclusions. The diagram indicates that $\Jac(C')$ is isomorphic to $\Prym(\tilde C,\upsigma)$.

Following \cite{Levin}, the curve $C'$ can be written in terms of $u$ and the symmetric functions $w+\bar w, \, w \bar w$.
In view of \eqref{big_C}, we have
$$
 (w+\bar w)^2 = 2 h(u) + 2 w \,\bar w, \quad   (w \, \bar w )^2 = h^2(u)- v^2 g^2(u) .
$$
Setting here $W= (w+ \bar w)/\sqrt{2}, Z= w \, \bar w$, one obtains equations defining $C'$ in
${\mathbb C}^3(u,Z,W)$:
\begin{equation} \label{2nd}
 C'\, :\; W^2 = h(u) + Z, \quad Z^2 = h^2(u)- g^2(u) \rP_6(u) ;
\end{equation}
this leads to the single equation $ C'\, :\; W^4 -2 h(u) W^2 +g^2(u) \rP_6(u)=0$. 

\begin{rem} 
In the  special case that the polynomial $h^2(u)- g^2(u) \rP_6(u)$ is a perfect square $f^2(u)$, the 
latter equation admits
the factorization
\begin{equation} \label{split}
      ( W^2- h(u)-f(u) ) (W^2- h(u)+f(u)) =0, 
\end{equation}
and $C'$ is a union of two curves whose regularizations give elliptic curves.  
This situation will be considered in detail elsewhere.
\end{rem} 
 
\paragraph{A hyperelliptic form of $C'$.} We return to the general case when 
$h^2(u)- g^2(u) \rP_6(u)$ is not a perfect square. 
Using the factorization \eqref{disc}, in \eqref{2nd} we can write
\begin{gather}
Z^2 = 16\,\mu ^{2}\,u^{6}\, (1 + \lambda \,u)^{2} \, t^2 \, (u-u_P)^2 F_1^2 \, Q^2(u),
\end{gather}
where we set
\begin{gather}
t^2 : = - \frac{\bar F_1}{F_1} \left(\frac{u-u_Q}{u-u_P}\right) = \frac{-\bar F_1 u+ \bar F_2}{F_1 u + F_2} \label{tt2}
\end{gather}
and $Q(u)$ is specified in \eqref{disc}.
Solving the last equation with respect to $u$, we get
\begin{equation} \label{ut}
u = \frac{\bar F_2 - t^2 F_2}{t^2 F_1+ \bar F_1}.
\end{equation}
Then equations \eqref{2nd} read
\begin{equation}
W^2 = h(u) + 4 \mu \, t\, u^3 \, (1 + \lambda \,u) \, Q(u)\, (F_1 u + F_2).   \label{hyper} 
\end{equation}
Replacing  $u$ here by \eqref{ut}, we obtain
\begin{gather}
 W^2 = 2 H  \frac{ {\cal P}^2(t) }{(t^2 F_1+ \bar F_1)^8 }(t+1)^2 \, \rR_6 (t) , \label{WR6} \\
 H = (\la+\mu)K_2 + \la^2-\mu^2 -K_1-1 = (\la+\mu) F_2- F_1 ,  \label{H}
\end{gather}
where ${\cal P}(t)$ is a polynomial in $t$ of degree 4, and $\rR_6(t)$ is specified below. 
We write 
$$
{\cal P}(t) = \frac{8 H^2}{(F_1 u+F_2 )^2} \Big( 2(\mu-\la) u^2+ (K_1-1) u+ F_2-2 \Big) -2 H t (t^2 F_1+\bar F_1),
$$
in a concise form, where $u$ should be replaced by \eqref{ut}.

Removing perfect squares from the right hand side of \eqref{WR6}, i.e. introducing a new variable $\cal W$ such that
\begin{equation} \label{W_W}
 W =\frac{ \sqrt{2H}\, {\cal P}(t) (t+1)  }{(t^2 F_1+ \bar F_1)^4 } \,    {\cal W},
\end{equation}
 we finally obtain $C'$ in hyperelliptic form as 
\beq \label{R6} 
   {\cal W}^2 = \rR_6(t) := c_6 t^6 + c_5 t^5 + \cdots + c_1 t +c_0 ,
\eeq 
$$ \bear{rl} 
\text{where}\,   & c_6= 
(\la+\mu) F_2^3 - (2\la-2\mu +1) F_1 F_2^2 +(\la+\mu+2) F_1^2 F_2 -F_1^3 , \\
 &c_0=               
(\la+\mu) {\bar F}_2^3 -(2\la-2\mu - 1) {\bar F}_1 {\bar F}_2^2 
+(\la+\mu-2){\bar F}_1^2 {\bar F}_2 + {\bar F}_1^3, \\
c_4  \, = & 
F_1^{3} -(\mu + \lambda+ 6)\,F_1^{2}\,F_2 -8(\mu+\lambda-1)\,F_1^{2} 
+ (6\lambda - 6\mu + 1)\,F_1\,F_2^{2} \\
& 
+ 8 (\mu^{2}+ 2\,\mu \lambda  + \lambda^{2} +4\,\mu - 1)\,F_2\,F_1 -(\mu+\lambda )\, F_2^{3}
 + 8(\mu^{2} - \lambda^{2} + \mu + \lambda ) \, F_2^{2} , \\
c_2  \, = & 
- {\bar F}_1^{3} - (\mu + \lambda - 6)\,{\bar F}_2\, {\bar F}_1^{2}
 + 8(\mu+ \lambda+ 1)\,{\bar F}_1^{2} + (6\,\lambda - 6\,\mu  - 1)\, {\bar F}_1\, {\bar F}_2^{2} \\
& 
+ 8(\mu ^{2} + 2\,\mu \lambda  + \lambda ^{2} - 
4\,\mu  - 1)\,{\bar F}_1\,{\bar F}_2 - (\mu+\lambda )\,{\bar F}_2^{3} 
+  8(\mu^{2} - \lambda^{2} - \mu - \lambda) \,{\bar F}_2^{2} , \\
& c_5 = 
2 H (F_1^2-F_2^2), \quad c_3 = 4H (F_1 \bar F_1 + F_2 \bar F_2 ), \quad 
c_1 = 2 H ({\bar F}_1^2-{\bar F}_2^2) .
\eear
$$
Equivalently, the equation (\ref{R6}) 
can be written compactly by using both variables $t$ and $u$, related by \eqref{ut}:
\beq  \label{short_C} 
\bear{rr}
\dfrac{(F_2 + F_1 u)^3 {\cal W}^2}{32H^3} = & (K_1+1) u^3  +(3\mu-3\la+ K_2) u^2+(K_1-3)u \\
&\qquad + \,\la -\mu + K_2 -t (F_2 + F_1 u ) (u^2-1) . \eear 
\eeq 

We can summarize the results of this section with
\begin{theorem} \label{Prym_to_Jac}
The Jacobian of the spectral curve $S$ in \eqref{spectral} contains a 2-dimensional Prym variety, isomorphic
to the Jacobian of the genus 2 curve $C'$ given by \eqref{R6} or \eqref{short_C}.
\end{theorem}

It is also worth mentioning the following relation between the roots of $\rR_6(t)$ and of the polynomial $\rP_6(u)$ defining
the first genus 2 curve $C=S/\upsigma$.
\begin{proposition} \label{roots_6}
If $t=\hat t$ is a root of $\rR_6(t)$ then \eqref{ut} gives a root of  $\rP_6(u)$.  
\end{proposition} 

 \begin{prf} 
For $t=\hat t$ we have ${\cal W}=0$, which, in view of \eqref{W_W}, implies $W=0$ (provided
that the denominator in \eqref{W_W} does not vanish for $t=\hat t$, and this condition always holds). In view of
the definition $W= (w+ \bar w)/\sqrt{2}$, in this case $w = -\bar w$, which by \eqref{big_C} gives $g(u)\, v=0$.
Next, since $g(u), \rP_6(u)$ do not have common roots and deg $g(u)=5$, the last equation defines 11 values of $u$, which,
via \eqref{tt2}, correspond to 11 zeros of the right hand side of \eqref{W_W}. Further calculations show that  
${\cal P}(t)\, (t+1)=0$ implies $g(u)=0$. As a result, the 6 zeros of ${\cal W}$ correspond to the 6 zeros of $v$, i.e.,
the roots of $\rP_6(u)$. 
\end{prf} 

\section{Translation on  $\Prym (S,\upsigma)$ and on $\Jac(C')$}
\setcounter{equation}{0}

Below we represent the reduced Somos 6 map $\hat\varphi \, :\; {\mathbb C}^4 \to {\mathbb C}^4$ 
as a translation on the Jacobian of the spectral curve $S$, given by a divisor  $\cal V$, 
and show that it belongs to $\Prym (S,\upsigma)$. 
Then the translation will be described in terms of degree zero divisors on the curve $C'$. 

First, recall that the Jacobian variety of an algebraic curve $X$ can be defined
as the additive group of degree zero divisors on $X$ considered modulo divisors of meromorphic functions on $X$.
Equivalence of divisors ${\cal D}_1, {\cal D}_2$ will be denoted as ${\cal D}_1 \equiv {\cal D}_2$.

Let ${\cal J}_K$ be the isospectral manifold, the set of all the matrices $\lax (x)$ of the form \eqref{Lax3} having the
same spectral curve $S$. Consider  the {\it eigenvector map}
$$
{\cal E}\, :\;  {\cal J}_{K} \longrightarrow \Jac(S) ,
$$
defined as follows: 
a matrix $\lax (x)\in {\cal J}_K$ induces
the eigenvector bundle ${\mathbb P}^2\to S$; for any point $p=(x,y)\in S$
$$
p \longrightarrow \psi (p) = (\psi_1(p), \psi_2(p), \psi_3(p))^T \quad \text{such that} \quad
\lax (x) \psi(p) = y\, \psi(p) .
$$
We assume that the eigenvector $\psi(p)$ is normalized: $\langle \upalpha, \psi(p)\rangle = 1$, for
a certain non-zero $\upalpha \in \C^3$. 
This defines the divisor $\cal D$ of poles of $\psi (p)$ on $S$.
For any choice of normalization, such divisors form an equivalence class $\{ {\cal D} \}$. The latter 
defines a point  ${\cal D}- N p_0 \in \Jac(S)$ with a certain base point $p_0\in S$. 
Here $N=\text{degree}({\cal D})$, and for  the case at hand $N=6$. Then 
${\cal E}(\lax (x)) = {\cal D}- 6 p_0$.      

Now let $\cal G$ be the maximal subgroup of ${\mathbb P}GL(3,{\mathbb C})$ which acts freely on ${\cal J}_{K}$ by conjugations and preserves the structure of  
$\lax (x)$. For any $g\in {\cal G}$ the $\cal E$-images of  $\lax (x)$ and $\hat \lax (x)= g \lax (x) g^{-1}$ give equivalent divisors. 
As was shown in e.g., \cite{AvM, Beau}, the reduced eigenvector map ${\cal E}' \, : \; {\cal J}_{K}/{\cal G} \mapsto \Jac(S)$ is injective.  
Note that, due to the specific structure of \eqref{Lax3},  in our case the subgroup $\cal G$ is trivial, and ${\cal J}_{K}/{\cal G} \cong {\cal J}_{K}$. 


\begin{theorem} \label{trans}
Under the map $\cal E$, the transformation ${\bf L}(x)\to \tilde {\bf L}(x)$ defined
by the discrete Lax representation \eqref{Lax3} is the translation on $\Jac (S)$ given by the degree zero divisor
$$
{\cal V}= \bar{\cal O} - {\cal O},
$$
where, as above, ${\cal O}=(x=0,y=0), \, \bar {\cal O}=(x=\infty,y=\infty)\in S$.
\end{theorem}
\begin{prf}
As follows from the intertwining relation \eqref{dlax},
if $\psi(p)$ is a normalized eigenvector of ${\bf L}(x)$, then
$\widehat \psi (p) = {\bf M}(x)\psi(p)$
is an eigenvector of $\tilde {\bf L}(x)$ with the same eigenvalue.
Note that, in contrast to $\psi(p)$,  
$\widehat \psi(p)$ is not normalized, as all of it components may vanish at some points of $S$, so we consider its normalization
$\tilde\psi(p)=f(p)^{-1} \widehat\psi(p)$, $f(p)=\langle \upalpha, \widehat\psi(p)\rangle$ for a generic non-zero normalization vector $\upalpha\in\C^3$. 

Compare the divisors $\cal D, \tilde{\cal D}$ of poles of $\psi (p), \tilde \psi (p)$.
Using (\ref{rpmap}) (or (\ref{mods6map}) in the appendix) we have 
$$
\det {\bf M}(x) =  \dfrac{R_1 R_2}{R_0^2}\, x,
$$ 
which implies that  ${\bf M}(x), {\bf M}^{-1}(x)$ are non-degenerate
apart from the points of $S$ over $x=0$ and $x=\infty$.
Then $\cal D, \tilde{\cal D}$ can differ only by the points
${\cal O}=(0,0), {\cal O}_1=(0,-\mu/\la)$ and $\bar {\cal O}=(\infty,\infty), \bar {\cal O}_1=(\infty,-\la/\mu)$ or their
multiples.

According to the structure of the matrix in \eqref{Lax3},
${\bf M}(0)$ has a eigenvalue 0 with multiplicity 2, 
with a 1-dimensional eigenspace spanned by the vector $(1,1,0)^T$, and
$\psi({\cal O})$ is parallel to $(1,1,0)^T$, whereas $\psi({\cal O}_1)$ is not.
That is,
$$
\widehat \psi ({\cal O})= {\bf M}(0) \psi({\cal O})=0, \quad \widehat \psi ({\cal O}_1)= {\bf M}(0) \psi({\cal O}_1)\ne 0.
$$
Further, for $\tau=\sqrt{x}$ being a local parameter on $S$ near $\cal O$, we have the expansion
$\psi(p)=( 1+O(\tau), 1+O(\tau), O(\tau))^T$. Then near $\cal O$,
$\widehat\psi(p) =(O(\tau), O(\tau), O(\tau))^T$, hence the normalizing factor
$f=\langle \upalpha, \widehat\psi(p)\rangle$ has a {\it simple} zero at $\cal O$ and does not vanish at
${\cal O}_1$.

Similarly, by considering the expansions of $\psi(p)$ near the points $\bar {\cal O}, \bar {\cal O}_1$ 
at  infinity,
one observes that $f(p)$ has a simple pole at $\bar {\cal O}$ and no poles at $\bar {\cal O}_1$.
As a result,
$$
   (f) = {\cal O} + {\cal U} - {\cal D} - \bar {\cal O} ,
$$
for a certain effective divisor ${\cal U}$. Then the divisor of poles of
$\tilde\psi(p)=f(p)^{-1} \widehat\psi(p)$ equals ${\cal U}$. Indeed, the zeros of
$\widehat\psi(p)$ and $f(p)$ at $\cal O$, as well as their poles
at $\mathcal D+ \bar{\cal O}$ cancel each other. Since $f(p)$ is meromorphic on $S$,
we conclude that $\tilde{\cal D}$ is equivalent to
${\cal D} + \bar {\cal O} - {\cal O}$. 
Thus the images of $\cal D$, $\tilde{\cal D}$ in
$\Jac(S)$ differ by the translation $\mathcal V$. 
\end{prf} 

Clearly $\upsigma({\cal V})= - {\cal V}$, hence the divisor ${\cal V}$ represents a vector on
$\Prym( S,\upsigma)$. Then, under the map $\cal E$, any orbit obtained by iterations of ${\bf L}(x)\to \tilde {\bf L}(x)$ belongs to a translate of $\Prym( S,\upsigma)\subset \Jac(S)$. 

Now observe that in our case the manifold  ${\cal J}_K$ coincides with ${\cal I}_K$, which has dimension 2. 
Then, since $\cal E$ is injective, ${\cal I}_K$ must be an open subset of $\Prym(S,\upsigma)$ or of a union of different translates of it. 
Note, however, that the latter is not a connected complex manifold, whereas, as was mentioned in Remark 5, ${\cal I}_K$ is an irreducible complex algebraic manifold, hence a connected one.   
Therefore, in view of Theorem \ref{Prym_to_Jac}, we arrive at the following result.  

\begin{theorem} \label{IK}
A generic complex invariant manifold ${\cal I}_K$ of the map $\hat\varphi$ is
isomorphic to an open subset of the Jacobian of the genus 2 curve $C'$ given by \eqref{R6} or \eqref{short_C}.
\end{theorem}

Upon comparing this result with the solutions \eqref{x_wp}  of  $\hat\varphi$ and with the properties of the embedding $\uppsi:\,\Jac(X)\setminus (\sigma_{0123})\to \mathbb{C}^4$,
 in the sequel it is natural to choose the genus 2 curve $X$ in \eqref{can_X} to be birationally equivalent to $C'$. 
\medskip

Now observe that for the special points ${\cal O}_1, \bar{\cal O}_1, {\cal O}_2 ,\bar {\cal O}_2 \in S$ 
described in Section 2, the 
degree zero divisors ${\cal V}_1={\cal O}- {\cal O}_1, {\cal V}_2={\cal O}_2-\bar {\cal O}_2$ are also antisymmetric
with respect to the involution $\upsigma$, 
hence they represent vectors in $\Prym(S,\upsigma)$ and, therefore, in $\Jac(C')$. 
Our next objective is to describe ${\cal V},{\cal V}_{1}, {\cal V}_{2}$ in terms of degree zero divisors on $C'$.  

\begin{theorem} \label{trans1}
Under the transformation of $S$ to the canonical form $\tilde C$ given by \eqref{tC}, the points ${\cal O}, \bar{\cal O}$
become
\begin{gather} \label{R12}
  {\cal R}_1 = (u=0, v=-1, w= 2),  \quad {\cal R}_2 = (u=0, v=-1, w=-2) .
\end{gather}
The points ${\cal O}_1, \bar{\cal O}_1\in S$
become ${\cal R}_{1}', {\cal R}_2' \in {\tilde C}$, the two preimages
of the infinite point $\infty_2\in \{\infty_1, \infty_2 \}\subset C$ specified by the
Laurent expansions $u=1/\tau$, $v= -(\la+\mu)/\tau^3 + O(\tau^{-2})$ with the local parameter $\tau=1/u$.
The expansions of  
$w$ near ${\cal R}_{1}', {\cal R}_2'$ are 
$$
w = \pm \frac{2 (\la^2-\mu^2)}{\tau^4}+ O(\tau^{-3}),
$$
respectively. 
Also, ${\cal O}_2, \bar{\cal O}_2$ become the points
\beq 
  {\cal R}_{1,2}''= \left( u=-\frac{1}{\la}, \, v= \frac{ \varkappa}{\la^3}, \,
w= \pm \, 2\mu \frac{\varkappa}{\la^4} \right ), 
 \label{R''}  
\quad  \varkappa = \la^3+ K_2\la^2 -(K_1+1) \la +\mu . 
\eeq 
\end{theorem}

\begin{prf} 
Relations \eqref{TY} describing the projection $\pi \,: S\to G$ imply that the coordinate $Y$ has poles at
$\pi({\cal O}), \pi(\bar{\cal O})\in G$. More precisely, in view of the behavior \eqref{div_x_y}, these are triple poles.
On the other hand, as follows from \eqref{TY}, \eqref{YY},
on the curve $C$ the function $Y(u,v)$ has a triple pole only at $(u=0,v=-1)$.
(In particular, at $(u=0, v=1)$ this function has a pole of lower order, and it has no poles at the two points 
at  infinity on $C$.)
Hence,  ${\cal O}, \bar{\cal O} \in S$ are projected to the same point $(u=0,v=-1)$ on $C$. Since it is not
a branch point of $\tilde C\to C$, the point has two preimages on $\tilde C$. To find their $w$-coordinates, we note
that \eqref{YY} implies
\begin{equation}
   \la\, \mu u^3(1+\la u)\, Y(u,v) \bigg |_{u=0, v=-1} =2, \quad
\la\, \mu u^3(1+\la u)\, Y(u,v) \bigg |_{u=0, v=1} =0,
\end{equation}
and for $u=0,v=-1$ the equation \eqref{tC} gives $w^2=4$, thus we get \eqref{R12}.
The proof of the rest of the theorem goes along similar lines. 
\end{prf} 

Now 
recall  \cite{Mum} that 
for a double cover of curves $\pi \, : \, C_2 \to C_1$ with an involution
$\upsigma\, : \, C_2 \to C_2$, there are two natural maps between $\Jac(C_2)$ and $\Jac(C_1) \subset \Jac(C_2)$:
\medskip

$\bullet$ the pullback $\pi^*  : \, \Jac(C_1) \to \Jac(C_2)$;  \\
$$
\text{degree 0 divisor $D$ on $C_1$} \: 
\to \;
\text{degree 0 divisor $\tilde D= {\pi}^{-1}(D)$ on $C_2$};
$$

$\bullet$ The projection ({\it Norm map}) $\Nm_{C_1} : \, \Jac(C_2) \to \Jac(C_1)$; \\ 
$$
\text{degree 0 divisor $\tilde D$ on $C_2$} \to \text{degree 0 divisor $\pi (\tilde D)$ on $C_1$}.
$$
Notice that for any degree zero  divisor $D$ on $C_1$, $\Nm_{C_1}(\pi^*(C_1))= 2\,C_1$. This property should be understood 
on the level of equivalence classes of divisors, that is,  
for a degree zero divisor $D$ on $C_1$, let $ {\tilde D}' $ be any divisor on $C_2$ equivalent to
$\tilde D={\pi}^{-1}(D)$. Then $\pi ( {\tilde D}') \equiv 2 D$ on $C_1$. 
Then let $ {\cal A}$, $\tilde{\cal A}$ be the Abel maps with images in $\Jac(C_1)$, $\Jac(C_2)$ respectively.
Hence
\begin{equation} \label{norms}
 \tilde {\cal A} ( {\tilde D}) = \frac 12 \tilde {\cal A} (\pi^* \circ \Nm_{C_1} ({\tilde D}' ) )\quad \text{and} \quad
{\cal A} (\Nm_{C_1} {\tilde D}') = 2\, {\cal A} (D).
\end{equation}

Now apply the above to the tower \eqref{tower}, introducing the map
$$
\Nm_{C'}\, : \; \Jac(\widetilde{\widetilde C}) \to \Jac(C') \subset \Jac(\widetilde{\widetilde C})
$$
as follows:
for a degree zero divisor $\cal Q$ on $\widetilde{\widetilde C}$, $\Nm_{C'} ( {\cal Q})=\pi_1 ({\cal Q})$.
Next, consider the sequence of divisors
\begin{equation}\label{seq1}
\begin{CD}  {\cal V}= \underbrace{ {\cal R}_1 - {\cal R}_2 }_{\in\, \Prym(\tilde C,\sigma)} @ > \tilde \pi^{-1} >>
 \tilde R_1^+ +\tilde R_1^- - \tilde R_2^+ -\tilde R_2^- @
> \pi_1 >>  {\cal S}= \underbrace{ S_1^+ + S_1^- - S_2^+ -S_2^- }_{\in\, \Jac(C')} ,
\end{CD}
\end{equation}
where $\tilde R_j^\pm$ are the preimages of ${\cal R}_j$ on $\widetilde{\widetilde C}$.

Let now $\cal A$ be the Abel map to $\Jac(C')$ and ${\cal V}_0$ be a degree zero divisor on $C'$ such that
the pullbacks $\tilde\pi^{-1}( {\cal R}_1 - {\cal R}_2 )$ and
$\pi_1^{-1}({\cal V}_0)$ give equivalent divisors on $\widetilde{\widetilde C}$. 
That is, the vector
$ {\bf w} ={\cal A}({\cal V}_0) \in \Jac(C')\subset \Jac(\widetilde{\widetilde C}) $ coincides with the Abel image
of ${\cal R}_1 - {\cal R}_2$ in $\Prym(\tilde C, \upsigma) \subset \Jac(\widetilde{\widetilde C})$.
Then, in view of the property \eqref{norms}, 
\begin{equation} \label{2S}
   {\bf w} = \frac 12 {\cal A} (S_1^+ + S_1^- - S_2^+ -S_2^- ).
\end{equation}

\begin{proposition} \label{SW} On the curve $C'$ written in the hyperelliptic form \eqref{R6} the divisor ${\cal V}_0$ is
determined by 
\begin{gather*}
S_1^+ = (t_*, {\cal W}_*^+) , \quad S_1^-= (-t_*, {\cal W}_*^-), \quad
S_2^+ = (t_*, -{\cal W}_*^+), \quad S_2^- = (-t_*, -{\cal W}_*^-) , \\
t_* = \sqrt{\frac{\bar F_2}{F_2}}, \quad {\cal W}_*^\pm = 4 \frac{H^{3/2}}{F_2} (1 \pm t_*) =
 4 \frac{H^{3/2}}{F_2^{3/2}} ( \sqrt{F_2} \pm \sqrt{\bar F_2}) .
\end{gather*}
\end{proposition}

The proof of the proposition is quite technical and is reserved for the appendix. 
Note that the squares of ${\cal W}_*^\pm$ given above
coincide with the right-hand side of \eqref{short_C} for $(u=0, t=\pm t_*)$, as expected.

One now can observe that the divisor ${\cal S}= S_1^+ + S_1^- - S_2^+ -S_2^-$ is anti-invariant with respect to
the hyperelliptic involution $\iota\, : \, (t,{\cal W}) \to (t,-{\cal W})$ on $C'$. Hence, modulo period vectors of
$\Jac (C')$,
$$
{\cal A}(S_1^+ - S_2^+ ) = 2 {\cal A}(S_1^+ - P_0) , \quad {\cal A}(S_1^- - S_2^-) = 2 {\cal A}(S_1^- - P_0),
$$
where $P_0$ is any Weierstrass point on $C'$. Then, using 
the relation \eqref{2S}, we arrive at

\begin{theorem} \label{Main}
The reduced Somos 6 map $\hat\varphi$ is described by translation by the following vector on $\Jac(C')$:
 \begin{equation} \label{vt}
 {\bf w}= \int_{P_0}^{ (t_*, {\cal W}_*^+) } (\omega_1, \omega_2)^T +
\int_{P_0}^{  (-t_*, {\cal W}_*^-) } (\omega_1, \omega_2)^T =
\int_{(-t_*, -{\cal W}_*^-) }^{(t_*, {\cal W}_*^+) }  (\omega_1, \omega_2)^T \, ,
\end{equation}
where $\omega_{1,2}$ are holomorphic differentials on $C'$
and $t_*, {\cal W}_{*}^\pm$ are given in Proposition \ref{SW}.
\end{theorem}

Above we have chosen the genus 2 curve $X$ of the form \eqref{can_X} to be birationally equivalent to $C'$,
and in the next section we shall see that the translation vector ${\bf v}\in \Jac(X)$ in the sigma-function solution \eqref{sigma_sol}
corresponds to the vector $\bf w$ written in an appropriate basis of holomorphic differentials on $C'$. Namely, it satisfies the constraint \eqref{det3}.
To show this it is convenient to describe also the degree zero divisors 
${\cal V}_1={\cal O}- {\cal O}_1, {\cal V}_2={\cal O}_2-\bar {\cal O}_2$ on $S$, 
or equivalently,  
${\cal R}_1' - {\cal R}_2', {\cal R}_1'' - {\cal R}_2''$ on $\tilde C$ (which are antisymmetric under $\upsigma$), 
in terms of divisors on $\Jac(C')$.
Namely,
let ${\cal V}_{01}, {\cal V}_{02}$ be degree zero divisors on $C'$ such that the vectors
$$
{\bf w}_1 ={\cal A}({\cal V}_{01}), \qquad 
{\bf w}_2 ={\cal A}({\cal V}_{02}) \in \Jac(C')\subset \Jac(\widetilde{\widetilde C})
$$
coincide with the Abel images of ${\cal R}_1' - {\cal R}_2'$, ${\cal R}_1'' - {\cal R}_2''$  respectively, 
in $\Prym(\tilde C, \upsigma) \subset \Jac(\widetilde{\widetilde C})$.

\begin{theorem} \label{VV12}
The divisors ${\cal V}_{01}, {\cal V}_{02}$ are equivalent to
$$
   (t_*', {\cal W}_*^{+'}) - (-t_*', -{\cal W}_*^{-'} ) 
\quad\text{and} \quad 
(t_*'', {\cal W}_*^{+''} ) - (-t_*'', - {\cal W}_*^{-''} ), 
$$
respectively, that is,
$$
  {\bf w}_1 = \int_{ (-t_*', -{\cal W}_*^{-'}) }^{ (t_*', {\cal W}_*^{+'}) }  (\omega_1, \omega_2)^t \, , \quad
 {\bf w}_2 = \int_{(-t_*'', -{\cal W}_*^{-''}) }^{(t_*'', {\cal W}_*^{+''}) }  (\omega_1, \omega_2)^t \, ,
$$
where
\begin{equation} \label{t''}
\begin{aligned}
   t_*' = \sqrt{\frac{-\bar F_1}{F_1}}, \quad {\cal W}_*^{\pm'} & = 4 \frac{H^{3/2}}{F_1} (1 \pm t_*') =
 4 \frac{H^{3/2}}{F_1^{3/2}} \left( \sqrt{F_1} \pm \sqrt{- \bar F_1} \right) , \\
  t_*'' = \sqrt{\frac{ \la \bar F_2 + \bar F_1}{ \la F_2- F_1 }}, \quad
{\cal W}_*^{\pm ''} & = \frac{4H^{3/2}}{\la F_2- F_1 } \left( \la-1 \pm  (\la+1) t_*''\right) \\
 &= \frac{4 H^{3/2}
\left( (\la+1)\sqrt{\la F_2- F_1} \pm (\la-1) \sqrt{\la \bar F_2 + \bar F_1} \right ) }{ ( \la F_2- F_1 )^{3/2}}.
\end{aligned}
\end{equation}
\end{theorem}

The proof follows the same lines as that of Theorem \ref{Main}; it uses Theorem \ref{trans1} describing
the coordinates of the 
pairs ${\cal R}_{1,2}', {\cal R}_{1,2}''$ on $\tilde C$. For each pair we construct
a sequence of degree zero divisors analogous to \eqref{seq1},
which gives rise to the divisors ${\cal V}_{01}, {\cal V}_{02}$ on $C'$ and, in view of the relation \eqref{2S},
the vectors ${\bf w}_1, {\bf w}_2$.

\section{The shift vector and the determinantal constraint}
\setcounter{equation}{0} 

The above vectors ${\bf w}, {\bf w}_1, {\bf w}_2\in \Jac(C')$
arising from the three special involutive pairs
${\cal O}, \bar{\cal O}, \, {\cal O}_1, \bar{\cal O}_1,\, {\cal O}_2,\bar{\cal O}_2$ on the spectral curve $S$
have the following remarkable property. 

\begin{proposition} The vectors in Theorem \ref{Main} and Theorem \ref{VV12} are related by 
\begin{equation} \label{123}
   {\bf w}_1 = - 2{\bf w}, \quad  {\bf w}_2 =- 3{\bf w} .
\end{equation}
\end{proposition}
%
\begin{prf} 
Following \eqref{div_x_y},  the divisor of the meromorphic function $y/x$ on $S$ is
\begin{equation} \label{div_y/x}
   (y/x) = {\cal O} +\bar {\cal O}_1 + {\cal O}_2 - \bar {\cal O} - {\cal O}_1 - \bar {\cal O}_2 \equiv 0 ,
\end{equation}
so it corresponds to zero in $\Jac(S)$, in $\Prym(S,\upsigma)=\Prym (\tilde C,\upsigma)$, 
and, therefore, in $\Jac(C')$.
By Theorems \ref{Main} and \ref{VV12},
the degree zero divisors ${\cal O}- \bar {\cal O}, {\cal O}_1 - \bar{\cal O}_1, {\cal O}_2-\bar {\cal O}_2$ with 
Abel image in $\Prym(S,\upsigma)$
are represented, respectively, by the divisors ${\cal V}_0 , {\cal V}_{01}, {\cal V}_{02}$ on $C'$. 
Then \eqref{div_y/x} implies ${\cal V}_0 - {\cal V}_{01} +{\cal V}_{02}\equiv 0$, which, under the Abel map, yields
${\bf w}- {\bf w}_1 + {\bf w}_2 = 0$.
Similarly, we have
$$
  (y/x^2) = \bar{\cal O} + 2 \bar {\cal O}_1 + {\cal O}_2 - {\cal O} - 2{\cal O}_1 - \bar {\cal O}_2 \equiv 0 ,
$$
which implies $-{\bf w} - 2{\bf w}_1 + {\bf w}_2 = 0$. These two relations prove the proposition. 
\end{prf} 

The divisors representing ${\bf w}_1, {\bf w}_2$ can also be derived (in a much more tedious way)
by using addition formulae on $\Jac(C')$ described
in terms of pairs of points on the hyperelliptic curve $C'$. These formulae can be obtained  algorithmically 
using  the B\"acklund transformation presented in \cite{Sab, KV}, which also allows us to calculate 
the divisor of the form 
$ (t_{+}''', {\cal W}_{+}^{'''}) - (t_{-}''', {\cal W}_{-}^{'''} )$ corresponding to the vector $4{\bf w}$.  
Since it will be needed in section 6, here we simply record that $t= t_{\pm}'''$ are the roots of the quadratic equation 
\begin{equation} \label{4wquad} 
(H(\lambda+1)+F_1)\,t^2 
+2\lambda H\,t+H(\lambda-1)+\bar{F}_1 = 0,   
\end{equation} 
and ${\cal W}_{+}^{'''}, {\cal W}_{-}^{'''}$ are recovered from the equation \eqref{R6} with their signs determined by the condition   
$$
{\cal W}_{+}^{'''} {\cal W}_{-}^{'''} =
64 
H^3 \frac{  
\mu ^{3} 
- \mu ^{2}\,\lambda ^{3} + \mu ^{2}\,\lambda ^{2}\,{K_2} 
- \mu ^{2}\,\lambda \,{K_1}
+ \mu
\,\lambda ^{3}\,{K_2} + \lambda ^{5} 
 }
{(H(\lambda+1)+F_1)^3\,
\left( 4 F_1 F_2 (\lambda F_2-F_1 )- H(F_1+F_2)^2 \right) } \,  . 
$$
\paragraph{The sigma-function, Kleinian functions, and the determinantal constraint.}
Now let an appropriate M\"{o}bius transformation $t={\cal T}(s)$ take the curve $C'$ to a canonical odd order form $X$
given by \eqref{can_X} (there are several possible transformations of this kind). Choosing  the canonical basis of
holomorphic differentials $\rd s/z, s\,\rd s/z$ 
on $X$, 
define the Abel map for a degree zero divisor $ (s_1,z_1) +(s_2,z_2)-2\infty$ by 
$$
 {\bf u} =(u_1, u_2)^T = \int_{\infty}^{(s_1,z_1) } \left( \frac{ds}{z}, \frac{s\, ds}{z} \right)^T +
\int_{\infty}^{ (s_2,z_2) } \left( \frac{ds}{z}, \frac{s\, ds}{z} \right)^T \, \in \Jac(X)\, .
$$
It can be inverted by means of the the Bolza formulae
\begin{gather*}
  (s-s_1)(s-s_2) = s^2- \wp_{22}({\bf u}) x - \wp_{12}({\bf u}), \\
 z_1 = \wp_{222}({\bf u}) s_1 + \wp_{122}({\bf u}), \quad
z_2 = \wp_{222}({\bf u}) s_2 + \wp_{122}({\bf u}) ,
\end{gather*}
which involve $\wp_{jk\ell}({\bf u} )= -\partial_j\partial_k\partial_\ell \log\sigma({\bf u})$  
in addition to  the Kleinian hyperelliptic functions $\wp_{jk}({\bf u})$. 
In particular, this yields
\begin{equation} \label{Bolza}
  s_1+ s_2 = \wp_{22}({\bf u}), \quad  s_1 s_2 = -\wp_{12}({\bf u}) , 
\end{equation}
and 
for $\wp_{11}$ there is also 
Klein's formula 
\beq\label{Klein} 
z_1z_2=\frac{1}{2}\sum_{k=0}^2 (s_1s_2)^k\Big(2\bar{c}_{2k}+(s_1+s_2)\bar{c}_{2k+1}\Big) 
-2(s_1-s_2)^2\wp_{11}({\bf u}).
\eeq 

Now let $(\bar s_1, \bar z_1), (\bar s_2, \bar z_2) \in X$ be the images of the points
$(t_*, {\cal W}_*^+), (-t_*, -{\cal W}_*^-) \in C'$ described in Proposition \ref{SW}, and
let 
$
\{ (\bar s_1', \bar z_1'), \, (\bar s_2', \bar z_2') \, \}, \; \{ \, (\bar s_1'', \bar z_1''),
(\bar s_2'', \bar z_2'') \, \} \in X\times X
$
be the 
images of
$\{ (t_*', {\cal W}_*^{+'}), (-t_*', -{\cal W}_*^{-'})\}, \, 
\{ (t_*'', {\cal W}_*^{+''}), (-t_*'', -{\cal W}_*^{-''}) \} \in C'\times C'$,  respectively, as specified in \eqref{t''}.

\begin{theorem} \label{constraint_satisfied}
The vector
\beq\label{vee}
 {\bf v} = \int^{(\bar s_1, \bar z_1) }_{(\bar s_2, \bar z_2) }
\left( \frac{ds}{z}, \frac{s\, ds}{z} \right)^T \, \in \Jac(X)
\eeq 
satisfies the determinantal constraint \eqref{det3}.
\end{theorem}
\begin{cor}\label{almost} 
For given values of $K_1,K_2,\la,\mu$, take the associated genus 2 curve  $C'$ from  
(\ref{R6}), transform it into the canonical form $X$, as in (\ref{can_X}), and pick the vector 
$\bv\in\Jac(X)$ defined by (\ref{vee}). 
Then for the function $\sigma({\bf u})$ associated with
$X$, and for any $\bv_0\in\Jac(X)$
and $\rC\in\C^*$, the expression \eqref{x_n} 
produces a  sequence $(\rx_n)$ satisfying the reduced Somos 6 recurrence \eqref{rec-4} with 
coefficients given by (\ref{abg}) and first integrals $K_1, K_2$.
\end{cor} 
\noindent{\bf Proof of Theorem \ref{constraint_satisfied}:} Observe that the vectors $\bf v$ and
$$
 {\bf v}_1 = \int^{(\bar s_1', \bar W_1) }_{(\bar s_2', \bar W_2) }
\left( \frac{ds}{W}, \frac{s\, ds}{W} \right)^T \, , \quad
{\bf v}_2 = \int^{(\bar s_1'', \bar W_1) }_{(\bar s_2'', \bar W_2) }
\left( \frac{ds}{W}, \frac{s\, ds}{W} \right)^T
$$
are just ${\bf w},{\bf w}_1, {\bf w}_2$ written in the coordinates corresponding to the canonical differentials on $X$.
Hence, the relations \eqref{123} imply ${\bf v}_1 = - 2{\bf v}, \; {\bf v}_2 =- 3{\bf v}$.
Then, in view of the Bolza expressions \eqref{Bolza}, the determinant constraint \eqref{det3} can be written as
 \begin{equation} \label{detc}
 \det \begin{pmatrix} 1 & 1 & 1 \\
  - \bar s_1 \bar s_2 & -\bar s_1' \bar s_2' & -\bar s_1'' \bar s_2''   \\
 \bar s_1+\bar s_2  &  \bar s_1'+ \bar s_2'   & \bar s_1''+ \bar s_2'' \end{pmatrix} =0 .
\end{equation}
Next, we apply the inverse M\"{o}bius transformation $s={\cal T}^{-1}(t)$ 
 to each of the above $s$-coordinates, and  
observe that, after dividing out common denominators from each column, the rows of the 
resulting 
matrix are linear combinations of the rows of 
\begin{equation} \label{detb}
 \begin{pmatrix} 1 & 1 & 1 \\
    (t_*)^2   &   (t_*')^2  &  (t_*'')^2    \\
 t_*-t_*  &  t_*' -t_*'   & t_*'' -t_*'' \end{pmatrix} .
\end{equation}
 As the third row of the latter matrix is zero, the condition \eqref{detc} is trivially satisfied.
\qed 

\begin{rem} 
To give an exact transformation from the equation of $C'$ to the
canonical form (\ref{can_X}) for $X$ one needs to know at least one root of the degree 6 polynomial
$\rR_6(t)$ in \eqref{R6} (or, in view of Proposition \ref{roots_6}, at least one root of $\rP_6(u)$ in \eqref{Gam}). 
Yet in  general it appears that the equation $\rR_6(t)=0$ is not solvable in radicals.  
\end{rem}


\section{Solution of  the initial value problem}
\setcounter{equation}{0} 

Before proceeding with the proof of Theorem \ref{main}, it is worth commenting on the meaning of the word ``generic'' 
appearing in its statement. Various  non-generic situations  arise:

\begin{itemize}
\item  Some of the initial data  or coefficients  can be zero.
\item For special values of $\la,\mu,K_1,K_2$, the spectral curve $S$  can acquire singularities (in addition to the singularity 
at $(0:1:0) \in{\mathbb P}^2$ for the projective curve). 
\item For special values of $\la,\mu,K_1,K_2$, the curve $C'$ becomes a  product of two elliptic curves given by 
the factorization (\ref{split}). 
\item  For special values of $\la,\mu,K_1,K_2$,  one of the multiples $2\bv,3\bv,4\bv$ of the shift vector 
can lie on the theta divisor $(\si )\subset\Jac(X)$. 
 \end{itemize} 

A set of non-zero initial data $\uptau_0,\ldots,\uptau_5$ and coefficients $\al,\be,\gam$ determine the values of 
$\la,\mu,K_1,K_2$, and these in turn determine the spectral curve $S$, the curve $C'$ (hence $X$), 
and the vector $\bv\in\Jac(X)$.
Yet  the sequence $(\uptau_n)$ may contain zero terms; for instance, $\uptau_0=0$ 
when $\bv_0={\bf 0}$ in (\ref{sigma_sol}).
Iteration of the recurrence  \eqref{S-6} requires non-zero initial data, but 
if an isolated  
zero appears in the sequence, then the Laurent phenomenon can be used to pass through this apparent singularity, 
by evaluating suitable Laurent polynomials in order to avoid 
division by zero. 

Another degenerate possibility is 
that one of $\al,\be,\gam$ 
is zero, in which case   \eqref{S-6}  can be obtained as a reduction of the Hirota-Miwa (discrete KP) equation, and the 
solutions require a separate treatment in each case. For each of  these three  special cases  there is also an associated 
cluster algebra, and by results in \cite{fockg,gsv} this means that a log-canonical symplectic structure is available for the reduced map (\ref{rec-4}) (see \cite{H10} for details).

If $\bv\in(\si)$, so that $\si(\bv)=0$, 
then  the expression (\ref{sigma_sol}) does not make sense; 
in that case one should replace  $\si(\bv)$ by $\si_2(\bv)=\partial_2\si(\bv)$ in the denominator of the formula for $\uptau_n$, and then it satisfies a Somos-8 recurrence \cite{beh}. This situation is not relevant to our 
construction, since it can be checked directly that for $\la,\mu,K_1,K_2$ such that (\ref{R6}) defines a curve of 
genus 2, neither $S_1^+$ nor $S_1^-$ in Proposition \ref{SW} can be a Weierstrass point on $C'$, 
hence $\bv\not\in (\si )$.
However,  it may happen that 
one of $2\bv,3\bv$ or $4\bv\in(\si )$, 
and in each  case the formulae in 
Theorem \ref{sigform} and/or our method for solving the initial value problem require certain adjustments. We illustrate 
this below in the case that $2\bv\in (\si)$, which is needed for reconstruction of the original Somos-6 sequence 
(\ref{classical}).

\paragraph{Reconstruction of the constant $\rC$ and the initial phase $\bv_0$.} 
The attentive reader might wonder why the constant $\rC$ in (\ref{sigma_sol}) 
should be necessary to represent the general solution of the Somos-6 recurrence. Indeed, making the scaling 
$s\to\zeta^2 \,s$, $z\to\zeta^5 z$  in  (\ref{can_X}) changes the coefficients $\bar{c}_j$ but preserves the form of the curve $X$, 
and rescales the sigma-function so that $\si ({\bf u})\to \zeta^{-3}\si ({\bf u})$, which means that 
$\rC$ can always be set to 1. 

Note, however, that the curve $C'$ equivalent to $X$ and given by \eqref{R6} is defined up to a similar rescaling 
$t\to \xi t, \, {\cal W} \to \xi^3 {\cal W}$, under which the vector $\bf w$ in Theorem \ref{Main} produces a family of points in $\Jac(C') \cong \Jac(X)$. The latter is precisely the curve 
in $\Jac(X)$ specified by the constraint (\ref{det3}).
Thus, to obtain the general solution of the recurrence, it is necessary to allow different values of $\rC$ in our  construction. 


Now if a particular set of non-zero initial data $\uptau_0,\ldots,\uptau_5$ and coefficients 
$\al,\be,\gam$ are given, then the associated initial values $\rx_0,\rx_1,\rx_2,\rx_3$ for the reduced map $\hat\varphi$
are found from (\ref{casi}). The latter values can be used in  
the formulae for $H_1,H_2$ in \cite{H10}, and from (\ref{h1h2}) these  produce the values of $K_1,K_2$; alternatively,  
putting the $\rx_j$ into \eqref{coord_PR} yields  $P_0,P_1, R_0, R_1$, and then $K_1,K_2$  can be obtained 
directly from the expressions  (\ref{ints_Kj}). The values of $\la$ and $\mu$ are specified according to (\ref{lamu}), and 
thus by Corollary \ref{almost} only $\rC$ and $\bv_0$ are needed to reconstruct the reduced sequence $(\rx_n)$. 

Supposing that $\bv,\bv_0$ and $\rC$ have already been found for a particular initial value problem, the parameters 
$\rA$ and $\rB$ are immediately obtained in the form 
\beq\label{AB} 
\rA = \frac{\rC\uptau_0}{ \sigma(\bv_0)} , \qquad   
\rB =  \frac{\sigma(\bv)\uptau_1}{ \rA \sigma(\bv_0+\bv)} 
  =  \frac{\sigma(\bv)\sigma(\bv_0)\uptau_1}{\rC \sigma(\bv_0+\bv) \uptau_0}. 
\eeq 
Thus the only outstanding problem is the determination of   $\rC$ and $\bv_0$. 

As an intermediate step, we introduce the sequence $(\tilde{\phi}_n)$ defined by 
\beq\label{companion}
\tilde{\phi}_n=\rC^{n^2-1}\, \frac{\si (n\bv)}{\si(\bv)^{n^2}}
\eeq 
Apart from the powers of $\rC$, the latter is the same as Kanayama's phi-function introduced in  \cite{Kan} in genus 2,
and  considered for hyperelliptic curves of arbitrary genus in \cite{uchida}. The sequence $(\tilde{\phi}_n)$ is a natural 
companion to $(\uptau_n)$: it satisfies the same Somos-6 recurrence \eqref{S-6} and produces the same values of the first integrals $K_1,K_2$. In fact, it turns out that for each $n$, $\tilde{\phi}_n$ is an algebraic function 
of the quantities $\al,\be,\gam,H_1,H_2$. (The proof will be presented elsewhere.) 
For our current purposes, it is enough to consider only the first  few terms of the sequence. 
\begin{lemma}\label{philem} The terms of the sequence  (\ref{companion}) for $n=0,\ldots,4$ are fixed up to signs by  
\beq\label{phis} 
\tilde{\phi}_0=0, \quad  
\tilde{\phi}_1=1, \quad 
\tilde{\phi}_2^2=\frac{\hat{\al}\be}{\al\hat\be}, \quad 
\tilde{\phi}_3^2=\frac{\beta}{\hat\be}, \quad 
\tilde{\phi}_4=\tilde{\phi}_2^{-1}(\al\tilde{\phi}_3-\gam).
\eeq 
\end{lemma} 
\begin{prf} For $n=0,1$ the result is immediate from the definition. The formulae for $n=2,3$ follow from   
 (\ref{abg}). To obtain $\tilde{\phi}_4$, set $n=-2$ in \eqref{S-6}, replace $\uptau_j$ with $\tilde{\phi}_j$ 
throughout, and use the fact that $\tilde{\phi}_{-j}=-\tilde{\phi}_j$ for all $j$. 
\end{prf} 

Now from the coordinates of the points on $C'$ considered in Proposition \ref{SW} and Theorem \ref{VV12}, 
and by applying the M\"{o}bius 
transformation $t={\cal T}(s)$ followed by the Bolza formulae together with (\ref{Klein}), 
as in the proof of Theorem 18, the values of the hyperelliptic functions 
$\wp_{jk}(m\bv)$ for $m=1,2,3$ are all determined algebraically in terms of $\la,\mu,K_1,K_2$. In turn, this allows 
$\hat\al$, $\hat\be$ to be found from (\ref{abhat}), which means that the expressions (\ref{phis}) determine  $\tilde{\phi}_2$, $\tilde{\phi}_3$ and 
$\tilde{\phi}_4$, up to fixing the signs of $\tilde{\phi}_2$, $\tilde{\phi}_3$. Then, upon rearranging the formula 
for $\gam$ in (\ref{abg}), we see that 
\beq\label{cform} 
\rC^2 =\frac{\gamma  }{\tilde{\phi}_3^2}
\Big( \wp_{11} (3 {\bf v} ) -\hat\alpha \wp_{11} (2 {\bf v} ) - \hat\beta  \wp_{11} ( {\bf v} ) \Big)^{-1}, 
\eeq 
which determines 
$\rC$ up to a choice of sign; and this sign is irrelevant, since from (\ref{AB}) the prefactor $\rA\rB^n\rC^{n^2-1}$ in 
(\ref{sigma_sol})  is seen to be invariant under sending $\rC\to -\rC$. Once $\rC$ is known, the sign of $\tilde{\phi}_3$
can then be fixed from an application of Baker's addition formula, which gives the identity 
$\tilde{\phi}_3=\tilde{\phi}_2^2\,{\cal F}(2\bv),$ 
where ${\cal F}$ is as in (\ref{x_wp}); the sign of $\tilde{\phi}_2$ will not be needed in what follows: it corresponds to the 
overall freedom to send $\bv\to -\bv,\bv_0\to -\bv_0$ in the solution, which is removed once the signs of the $z$ coordinates of the points in (\ref{vee}) are fixed.
\medskip

\noindent {\bf Proof of Theorem \ref{main}:}
Given the six non-zero initial data for  \eqref{S-6}, with the associated values of $K_1,K_2,\la,\mu$ being obtained as previously described, one finds the 
corresponding genus 2 curves $C', X$ and the vector $\bv\in\Jac(X)$. If $\rC$ is fixed from (\ref{cform}), then Theorem 1 says that for any $\rA$, $\rB$ and $\bv_0$ the expression (\ref{sigma_sol}), with this choice of $X$ and $\bv$, provides a solution of  \eqref{S-6} with the appropriate values of the coefficients $\alpha,\beta,\gamma$. To find the correct value of $\bv_0$, 
one should iterate the Somos-6 recurrence forwards/backwards to obtain additional 
terms, in order to calculate ratios of the form $\uptau_{j}\uptau_{-j}/\uptau_0^2$ for $j=1,2,3,4$. 
(By adjusting the offset of the index if necessary, a maximum of three iterations are needed to obtain 9 adjacent terms 
$\uptau_{-4},\uptau_{-3},\ldots,\uptau_4$ with generic initial data.) Matching these ratios with the analytic formula (\ref{sigma_sol}), 
and using Baker's addition formula,  
yields four linear equations for the quantities $\wp_{jk}({\bf v}_0)$, namely 
\beq \label{v0sys} 
\frac{\uptau_j\uptau_{-j}}{\uptau_0^2} = \rC^2 \tilde{\phi}_j^2\, \Big( \wp_{12}(j{\bf v})\wp_{22}({\bf v}_0) 
- \wp_{22}(j{\bf v})\wp_{12}({\bf v}_0) 
+ \wp_{11}(j{\bf v})-\wp_{11}({\bf v}_0)\Big)  
\eeq  
for $j=1,2,3,4$, 
Now, observing  that the first three equations are linearly dependent, due to the constraint  (\ref{det3}), it is necessary to 
use any two of the first three together with the fourth; for instance, picking $j=1,3,4$ produces the $3\times 3$ matrix equation  
\beq\label{134sys} 
 \left(\bear{ccc} 
\wp_{22}({\bf v}) & \wp_{12}({\bf v}) & 1 \\ 
\wp_{22} (3{\bf v})& \wp_{12}(3{\bf v}) & 1 \\ 
\wp_{22}(4{\bf v}) & \wp_{12}(4{\bf v}) & 1 
\eear\right) 
\left(\bear{c} -\wp_{12}({\bf v}_0)\\ \wp_{22}({\bf v}_0)\\ -\wp_{11}({\bf v}_0) 
 \eear\right)
=\left(\bear{c} \rC^{-2}\rho_1 - \wp_{11}({\bf v}) \\ 
\rC^{-2}\rho_3 - \wp_{11}(3{\bf v}) \\ 
\rC^{-2} \rho_4- \wp_{11}(4{\bf v})
\eear\right) ,  
\eeq 
where we set $$\rho_j = \uptau_j\uptau_{-j}/(\tilde{\phi}_j\uptau_0)^2.$$ 
In order to make this formula effective, the values of $\wp_{jk}(4\bv)$ are required; these can be found by taking 
the roots of (\ref{4wquad}) and transforming them with $s={\cal T}^{-1}(t)$  to the corresponding $s$-coordinates on $X$, 
or by directly applying the B\"acklund transformation for the genus 2 odd Mumford system \cite{Sab, KV} to perform the 
addition $3\bv+\bv=4\bv$ on $\Jac(X)$.
Upon solving (\ref{134sys}), the quantities $\wp_{jk}({\bf v}_0)$ are found, so that $\bv_0\in\Jac(X)$ is 
\beq\label{vee0} 
\bv_0 = \int_\infty^{(s^{(0)}_1,z^{(0)}_1)} \left(\frac{\rd s}{z},\frac{s\,\rd s}{z}\right)^T + \int_\infty^{(s^{(0)}_2,z^{(0)}_2)} \left(\frac{\rd s}{z},\frac{s\,\rd s}{z}\right)^T,
\eeq   
corresponding to the Abel map for the divisor ${\cal D}_0=(s^{(0)}_1,z^{(0)}_1)+(s^{(0)}_2,z^{(0)}_2)-2\infty$ on $X$, 
where the coordinates of the points $(s^{(0)}_1,z^{(0)}_1),(s^{(0)}_2,z^{(0)}_2)$ are obtained by using 
(\ref{Bolza}) and (\ref{Klein}) with ${\bf u}=\bv_0$. (An overall choice of sign for $z^{(0)}_j$ is left undetermined; this 
can be fixed by doing a single iteration, taking $\bv_0\to\bv_0+\bv$ and checking the result.)  Once $\bv_0$ has been found, the appropriate values of 
$\rA$ and $\rB$ are given by (\ref{AB}), and the initial value problem is solved. This completes the proof 
of Theorem \ref{main}. \qed
\medskip

We now show how Theorem \ref{Kum} is a corollary of this result.
\medskip 

\noindent{\bf Proof of  Theorem \ref{Kum}:} Given a point $(\rx_0,\rx_1,\rx_2,\rx_3)\in\C^4$ lying on a fixed invariant surface
${\cal I}_K =\{ K_1({\mathbf x})=k_1,  K_2({\mathbf x})=k_2\} $ of 
$\hat\varphi$, we can iterate the map forwards/backwards to obtain ratios of $\rx_j$ which correspond to the quantities on the 
left-hand side of (\ref{v0sys}), that is 
$$
\rx_{-1}=    
\frac{\uptau_1\uptau_{-1}}{\uptau_0^2}, \qquad \rx_{-2}\rx_{-1}^2\rx_0= 
\frac{\uptau_2\uptau_{-2}}{\uptau_0^2}, 
$$ 
and so on. This means that the initial vector $\bv_0$ is also recovered from a point on ${\cal I}_K$, which yields a vector ${\bf u} =\bv+\bv_0\in  \Jac(X)$, 
so the map (\ref{uppmap}) is invertible on each invariant surface, giving the required isomorphism on an open set of $\Jac(X)$ (removing the theta divisor and suitable translates).
\qed 
\medskip

\paragraph{A numerical example.}  
For illustration of the main result, we consider the following choice of initial data and coefficients:
$$ 
(\uptau_0,\ldots,\uptau_5) = (1,1,-1,1,-3,-3), \qquad \al=1, \qquad \be =2, \qquad \gam =-2.
$$ 
This produces an integer sequence which extends both backwards and forwards, 
\beq\label{exseq} 
\ldots,1,-1,1,1,1,1,-1,1,-3,-3,1,-25,49,1,385,1489,503,10753,-82371,\ldots, 
\eeq 
so that it has the symmetry $\uptau_n=\uptau_{-n-1}$. The corresponding 
initial data for \eqref{rec-4} are $(\rx_0,\rx_1,\rx_2,\rx_3)=(-1,1,3,-1/3)$, and so the 
first integrals presented in \cite{H10} take the values $H_1=-12,H_2=8$. Then (fixing a choice of square root) 
$\delta_1=1/2$, which gives 
\beq\label{params} 
\la = 2, \qquad \mu=-1, \qquad K_1=-6, \qquad K_2=-4 . 
\eeq 
After rescaling ${\cal W}$ suitably, the curve $C'$ is found from (\ref{R6}) to be 
\beq\label{cpex} 
C': \qquad {\cal W}^2=(t^2-1)(19t^4+16t^3+2t^2-80t-37). 
\eeq 
With the M\"{o}bius transformation $t={\cal T}(s)=(2s-5)/(2s+5)$ this is transformed to 
\beq\label{xex} 
X:\qquad z^2=4s^5 +52s^4+35s^3+25s^2-\frac{375}{4}s. 
\eeq 

Now to obtain the vector $\bv\in\Jac(X)$ as in Theorem \ref{constraint_satisfied}, we start from the points on $C'$
given in Proposition \ref{SW}, and applying the inverse  M\"{o}bius transformation 
$s={\cal T}^{-1}(t)$  to find the degree zero divisor $\cal D$ 
on $X$  corresponding 
to $\bv$. This produces
\beq\label{dform}
{\cal D}= (s_1,z_1)+(s_2,z_2)-2\infty,\qquad
\bv = \int_\infty^{(s_1,z_1)} \left(\frac{\rd s}{z},\frac{s\,\rd s}{z}\right)^T + \int_\infty^{(s_2,z_2)} \left(\frac{\rd s}{z},\frac{s\,\rd s}{z}\right)^T,
\eeq 
$$
\text{where}\qquad s_{1,2}= - \frac{5}{2}e^{\mp i\pi /3}, \qquad  z_{1,2}=25\sqrt{2}e^{\pm i\pi /3}.
$$ 
(Note that  we slightly changed notation here compared with  
(\ref{vee}); in particular, we dropped bars on the coordinates.) Similarly, by applying 
the same (inverse) M\"{o}bius transformation to the coordinates given in (\ref{t''}) and (\ref{4wquad}), or 
via the B\"acklund transformation in \cite{Sab, KV}, 
we find the divisors  corresponding to 
$2\bv,3\bv,4\bv\in\Jac(X)$, namely 
${\cal D}'= (s_1',z_1')+(s_2',z_2')-2\infty,{\cal D}''= (s_1'',z_1'')+(s_2'',z_2'')-2\infty, 
{\cal D}'''= (s_1''',z_1''')+(s_2''',z_2''')-2\infty$, where 
$$
s'_{1,2}= - \frac{5}{4} (5\pm\sqrt{21}), \,
z'_{1,2}=\frac{25}{2}\sqrt{2} (5\pm\sqrt{21}),\qquad 
s''_{1,2}=  \frac{5}{4} (3\pm\sqrt{5}), \, 
z''_{1,2}=\frac{25}{2}\sqrt{2} (11\pm 5\sqrt{5}), 
$$
$$ 
s'''_{1,2}=  \frac{5}{36} (-1\pm i\sqrt{35}), \qquad 
z'''_{1,2}=\frac{25}{486} \sqrt{2}(103\mp 13i\sqrt{35}).
$$  
Using the Bolza formulae (\ref{Bolza}) and (\ref{Klein}), this allows us to calculate the values of the 
Kleinian functions $\wp_{jk}(m\bv)$ for $m=1,2,3,4$, as presented in Table \ref{table:1}. 

\begin{table}[h!]
\centering
\begin{tabular}{||c | c| c |c||} 
 \hline
 $m$ & $\wp_{12}(m\bv )$ & $\wp_{22}(m\bv )$ & $\wp_{11}(m\bv )$ \\ [0.5ex] 
 \hline\hline
 1 & $-25/4$  &  $-5/2$   &  $-{125/8}$ \\ 
 2 &  $-25/4$ &  $-{25/2}$ &  $-{25/8}$ \\
 3 & $-25/4$  &  $ {15/2}$   &   ${575/8}$ \\
 4 & $-25/36$ &  $-{5/18}$  &   ${475/72}$\\ 
 \hline
\end{tabular}
\caption{Values of Kleinian functions at multiples of $\bv$ for sequence (\ref{exseq}). }
\label{table:1}
\end{table}

The values in the latter table, together with (\ref{abhat}), yield $\hat\al=-1$, $\hat\be = 2$, 
so from (\ref{phis}) we see that $\tilde{\phi}_2^2=-1$, $\tilde{\phi}_3^2=1$ and 
$\tilde{\phi}_4=\tilde{\phi}_2^{-1}(\tilde{\phi}_3+2)$. From (\ref{cform}), this is enough to determine that 
$\rC^2=-1/50$, and then from $\tilde{\phi}_3=\tilde{\phi}_2^2\,{\cal F}(2\bv)$ we find 
$$ 
\tilde{\phi}_2=\pm i, \qquad \tilde{\phi}_3=1, \qquad \tilde{\phi}_4=\mp 3i. 
$$ 
Then we have $\rho_1=1,\rho_3=-1,\rho_4=1/3$, which means that the linear system 
(\ref{134sys}) can be solved for $\wp_{jk}(\bv_0)$, to yield 
$$ 
 \rC=\frac{i}{\sqrt{50}},\qquad
\wp_{12}(\bv_0) = - 5/4 , \quad 
\wp_{22}(\bv_0) = 3/2 , \quad 
\wp_{11}(\bv_0) = 175/8 
$$
(where we have recorded a particular choice of sign for $\rC$). Hence, 
after fixing signs of the $z$-coordinates appropriately, 
the coordinates of the points in 
the divisor ${\cal D}_0$ are 
$$ 
s^{(0)}_{1,2}=
\frac{1}{4}(3\pm i\sqrt{11}), 
\qquad 
z^{(0)}_{1,2}=\sqrt{2}(1\mp 3i \sqrt{11}),  
$$
so that $\bv_0$ is given by (\ref{vee0}). Finally, with these values  of $\bv_0$ and $\bv$, 
the constants $\rA, \rB$ are found from (\ref{AB}) to be 
$\rA=i/(\sqrt{50}\,\si(\bv_0))$, $\rB=-i\sqrt{50}\,\si(\bv_0)\si(\bv)/\si(\bv_0+\bv)$.

\paragraph{The special case where $2\bv\in(\si)$.} In order to illustrate the modifications that are needed in a degenerate case, we briefly consider the situation where $2\bv$ lies on the theta divisor. This corresponds to having 
\beq\label{2v}
2\bv =\int_\infty^{(s',z')}  \left(\frac{\rd s}{z},\frac{s\,\rd s}{z}\right)^T,
\eeq 
the  image of a single point $(s',z')$ under the Abel map based at infinity. The formula (\ref{sigma_sol}) still makes sense, 
but (since $\wp_{jk}(2\bv )$ become singular) the expressions (\ref{abg}) for the coefficients are no longer 
appropriate, and  Theorem \ref{sigform} requires a slight reformulation. 
\begin{theorem}\label{modsigform} 
For $\bv\in\Jac(X)$ such that $2\bv$ has the form (\ref{2v}) modulo periods, with arbitrary $\rA,\rB,\rC\in {\mathbb C}^*,\, {\bf v}_0\in {\mathbb C}^2$, 
the sequence with $n$th term (\ref{sigma_sol}) 
satisfies the recurrence \eqref{S-6} with coefficients given by
\beq  \alpha = \frac{ \tilde{\phi}_3^2 }{\hat{\phi}_2^2} \, \hat\alpha , 
\qquad
\beta =\tilde{\phi}_3^2,  \qquad  
\gamma = \tilde{\phi}_3^2 \,\rC^2 
\Big( \wp_{11} (3 {\bf v} ) +\hat\alpha (s')^2 -   \wp_{11} ( {\bf v} ) \Big) , \label{nabg}
\eeq
where 
$\hat{\phi}_2=\rC^3\, \si_2(2\bv) /\si(\bv)^4$, 
$\si_2(u)=\partial_2 \si({\bf u})$, $ \hat\al = \wp_{22}(\bv)-\wp_{22}(3\bv)$, 
provided that $\bv$ satisfies the constraint 
$$ 
\det\begin{pmatrix} 1 & 0 & 1 \\
   \wp_{12} ( {\bf v} ) & -s' &  \wp_{12}(3 {\bf v} ) \\
 \wp_{22}({\bf v} ) & 1 &  \wp_{22}(3{\bf v} ) \end{pmatrix} =0 . 
$$
The coefficients satisfy the condition 
\beq\label{2vcon} 
\al^2 \be=\gam^2. 
\eeq 
\end{theorem} 
\begin{prf} The main formulae 
above 
arise from   Theorem \ref{sigform} by taking the 
limit $\si(2\bv)\to 0$ with $\si_2(2\bv)\neq 0$, or directly by using Baker's addition formula and its limiting case for 
a shift on the theta divisor \cite{beh}.
For the necessary condition (\ref{2vcon})  note that 
by \eqref{S-6} for $n=-2$ 
with $\uptau_j=\tilde{\phi}_j$ for all $j$ and $\tilde{\phi}_2=0$, 
the identity 
$\al\tilde{\phi}_3=\gam$ holds; squaring both sides of the latter 
and comparing with 
$\be$ in (\ref{nabg}) yields the condition. 
\end{prf} 

For the purpose of the reconstruction problem, we need an additional formula, namely  
\beq\label{comids} 
\tilde{\phi}_4^2=\al^3 +H_1. 
\eeq 
Its proof is based on the fact that   the  companion sequence also satisfies a Somos-10 recurrence (see Proposition 
2.5 in \cite{H10}), but we omit further details.  

\paragraph{The original Somos-6 sequence.} 
For the original sequence (\ref{classical}) considered by Somos, we choose to index the terms so 
that 
$$ 
(\uptau_0,\ldots,\uptau_5) = (1,1,1,3,5,9), \qquad \al=1, \qquad \be =1, \qquad \gam =1.
$$ 
We have $H_1=19,H_2=14$, as noted in \cite{H10}. Then (upon fixing a sign) 
$\delta_1=i$, giving
\beq\label{param2} 
\la = i, \qquad \mu=-i, \qquad K_1=19, \qquad K_2=14 i. 
\eeq 
The curve $C'$, found from (\ref{R6}), takes the form 
\beq\label{cpex2} 
C': \qquad {\cal W}^2=(t-1) \, \rQ(t), %
\eeq 
where, after removal of a numerical prefactor, $\rQ(t) =159025 t^5+\ldots + 154607+37224 i$ 
is a quintic polynomial with Gaussian integer coefficients whose real and imaginary parts have 5 or 6 digits.
The M\"{o}bius transformation $t={\cal T}(s)=(s-i)/(s+i)$ sends the root $t=1$ to infinity, and transforms 
$C'$ to the 
canonical quintic curve  
\beq\label{xex2} 
X:\qquad z^2=4 s^5 -233 s^4+1624 s^3-422 s^2+36 s-1. 
\eeq 

As in the previous example, by rewriting the points in Proposition \ref{SW} in terms of points in $X$, we obtain the divisor $\cal D$ and  corresponding vector $\bv\in\Jac(X)$ in the form (\ref{dform}), where 
$$
\qquad s_{1,2}= - 8\pm \sqrt{65}, \qquad  z_{1,2}=20i (129\mp16\sqrt{65}).
$$ 
The condition (\ref{2vcon}) clearly holds, but it is necessary, not sufficient for $2\bv\in (\si )$. 
However, from the first formula in (\ref{t''}) we find that $(t_*')^2=1$, meaning that one of the points in the divisor 
${\cal V}_{01}$ is the Weierstrass point $(1,0)\in C'$, and under the  M\"{o}bius transformation this means that 
$2\bv$ has the form (\ref{2v}) with $(s',z')=(0,-i)$, corresponding to the divisor 
$$ 
{\cal D}'=(0,-i)-\infty
$$ 
on $X$. Application of the formula for $t_*''$ in  (\ref{t''}) leads to the divisor 
${\cal D}''= (s_1'',z_1'')+(s_2'',z_2'')-2\infty$ corresponding to 
$3\bv$, where 
$$
s''_{1,2}=   -18\pm5\sqrt{13}, \qquad
z''_{1,2}=20i (-667\pm 185\sqrt{13}), 
$$
while for  
$4\bv$ we have ${\cal D}'''=2{\cal D}'$. 
This means that the finite values of 
$\wp_{jk}(m\bv)$ for $m=1,3,4$, as presented in Table \ref{table:2},  can be obtained in the usual way, except that (\ref{Klein}) is no longer 
valid for ${\bf u}=4\bv$. Instead, in order to find $\wp_{11}(4\bv)$, 
 we use the equation of the Kummer surface (see e.g. 
\cite{Baker2}), which provides a quartic relation between the functions  $\wp_{jk}({\bf u})$.

\begin{table}[h!]
\centering
\begin{tabular}{||c | c| c |c||} 
 \hline
 $j$ & $\wp_{12}(j\bv )$ & $\wp_{22}(j\bv )$ & $\wp_{11}(j\bv )$ \\ [0.5ex] 
 \hline\hline
 1 & $1$  &  $-16$   &  ${51/2}$ \\ 
 2 &  $\infty$ &  $\infty$ &  $\infty$ \\
 3 & $1$  &  $ -36$   &   ${11/2}$ \\
 4 & $0$ &  $0$  &   ${49/2}$\\ 
 \hline
\end{tabular}
\caption{Values of Kleinian functions at multiples of $\bv$ for sequence (\ref{classical}). }
\label{table:2}
\end{table}

From  the identities  in Theorem \ref{modsigform} and its proof we see that $\hat\al=20$ and $\tilde{\phi}_3=1$, giving 
$\hat{\phi}_2^2=20$, $\rC^2=-1/20$ from the first and last formulae in (\ref{nabg}), and also 
$\tilde{\phi}_4^2=20$ by (\ref{comids}).  
The equation (\ref{v0sys}) should be modified when $j=2$, but is valid for $j=1,3,4$, which means that 
we can still solve (\ref{134sys}) with $\rho_1=1,\rho_3=3,\rho_4=3/4$, and (fixing the sign of $\rC$) we find 
$$ 
\rC=\frac{i}{\sqrt{20}}, \qquad 
\wp_{12}(\bv_0) = - 1 , \quad 
\wp_{22}(\bv_0) = 10 , \quad 
\wp_{11}(\bv_0) = 79/2. 
$$
Hence 
$\bv_0$ is given by (\ref{vee0}), 
where 
the associated divisor ${\cal D}_0$ contains the coordinates
$$ 
s^{(0)}_{1,2}=5\pm 2\sqrt{6}, 
\qquad 
z^{(0)}_{1,2}=4i(71\pm 29 \sqrt{6}). 
$$
Thus, from (\ref{AB}), we see that 
$\rA=i/(\sqrt{20}\,\si(\bv_0))$, $\rB=-i\sqrt{20}\,\si(\bv_0)\si(\bv)/\si(\bv_0+\bv)$.

\section{Conclusion} 
\setcounter{equation}{0}  

The explicit solution (\ref{sigma_sol})  is equivalent to an expression 
$\uptau_n=A\, B^n\, C^{n^2}\,\Theta ({\bf z}_0+n{\bf z})$ in terms of a Riemann theta function in two variables, 
for suitable constants $A,B,C$ and vectors ${\bf z}_0,{\bf z}\in\C^2$. 
In fact (see {\tt http://somos.crg4.com/somos6.html}), 
a numerical fit of 
(\ref{classical}) with a two-variable Fourier series was performed by Somos some time ago. 

There are a number of aspects of the solution that we intend to consider in more detail elsewhere. The 
companion sequence (\ref{companion}) deserves more attention, since its properties should be helpful in proving 
that other Somos-6 sequences consist entirely of integers, in cases  where the Laurent property is insufficient; 
for example, take 
$$ 
(\uptau_0,\ldots,\uptau_5) = (2,3,6,18,54,108), \qquad \al=18, \qquad \be =36, \qquad \gam =108, 
$$ 
which defines a sequence belonging to an infinite family found by Melanie de Boeck. We also propose to examine 
Somos-6  sequences that are parametrized by elliptic functions, including the case where the factorization (\ref{split}) holds. 

The solution of the initial value problem for  \eqref{S-6}  raises the further possibility of performing 
separation of variables for  the reduced
map $\hat \varphi$ defined by (\ref{rec-4}), 
and finding a $2\times 2$ matrix Lax representation for it. In principle, such a representation  might also be used to obtain 
a symplectic structure for $\hat\varphi$ when $\al\be\gam\neq 0$, which could shed some light on the open problem 
of finding compatible Poisson or (pre)symplectic structures for general Laurent phenomenon algebras (see \cite{LP}).

\vspace{.1in} 
\noindent 
{\bf Acknowledgments:} The authors would like to thank the organisers of the WE-Heraus-Seminar on 
{\it Algebro-geometric Methods in Fundamental Physics}  in  Bad Honnef, Germany in September 2012, 
where we began to discuss this problem in detail. They are also 
grateful to Victor Enolski for interesting conversations on related matters and to the anonymous referee for remarks and suggestions.
 
YNF acknowledges support of the Spanish MINECO-FEDER Grants
MTM2015-65715-P, MTM2012-37070, and the Catalan Grant 2014SGR504.  
ANWH is supported by 
Fellowship EP/M004333/1 from the Engineering and Physical Sciences Research Council.

\subsection*{Appendix A. Derivation of the Lax pair}
The general Somos-6 recurrence arises by reduction from the discrete BKP equation, which is a given by a bilinear relation
for a tau function $T(n_1,n_2,n_3)$ that depends on three independent variables. For convenience we use indices to write    
$T(n_1,n_2,n_3) =T_{jk\ell}$,  $(n_1,n_2,n_3)=(j,k,\ell )$, 
so that the discrete BKP equation takes the form 
\beq\label{bkp} \bear{rcl} 
T_{j+1,k+1,\ell+1}T_{jk\ell}& - &  T_{j+1,k,\ell}T_{j,k+1,\ell+1}\\ 
&+ & T_{j,k+1,\ell}T_{j+1,k,\ell+1}-T_{j,k,\ell+1}T_{j+1,k+1,\ell}=0.
\eear 
\eeq
Following \cite{bkp}, this equation arises as a compatibility conditon  
of the linear triad  
\beq\label{triad} \bear{rcl} 
\Psi_{j+1,k+1,\ell}-\Psi_{jk\ell} & = & \dfrac{T_{j+1,k,\ell}T_{j,k+1,\ell}}{T_{j+1,k+1,\ell}T_{jk\ell}} 
\Big(\Psi_{j+1,k,\ell}-\Psi_{j,k+1,\ell}\Big), \\ 
\Psi_{j,k+1,\ell+1}-\Psi_{jk\ell} & = & \dfrac{T_{j,k+1,\ell}T_{j,k,\ell+1}}{T_{j,k+1,\ell+1}T_{jk\ell}} 
\Big(\Psi_{j,k+1,\ell}-\Psi_{j,k,\ell+1}\Big), \\
\Psi_{j+1,k,\ell+1}-\Psi_{jk\ell} & = & \dfrac{T_{j+1,k,\ell}T_{j,k,\ell+1}}{T_{j+1,k,\ell+1}T_{jk\ell}} 
\Big(\Psi_{j+1,k,\ell}-\Psi_{j,k,\ell+1}\Big).
\eear
\eeq 

Now construct a tau function 
which, apart from a gauge transformation by the exponential of a quadratic form, 
depends only on the single independent variable 
\beq\label{ndef} 
n=n_1+2n_2+3n_3=j+2k+3\ell; 
\eeq 
so we set 
\beq\label{taured} 
T_{jk\ell}=\delta_1^{k\ell} \delta_2^{j\ell} \delta_3^{jk}\, \uptau_n 
\eeq 
for some parameters $\delta_j$, $j=1,2,3$.
Substituting this into (\ref{bkp}) produces the Somos-6 recurrence for $\uptau_n$ with the coefficients   
\begin{equation} \label{deltas}
\alpha=\frac{1}{\delta_2\delta_3}, \quad \beta=-\frac{1}{\delta_1\delta_3}, 
\quad \gamma =\frac{1}{\delta_1\delta_2}.
\end{equation}

The derivation of the Lax pair for Somos-6 is somewhat more involved, 
but proceeds by applying an analogous reduction procedure to the system (\ref{triad}). 
We suppose that the wave function also depends primarily on the same dependent variable $n$ in (\ref{ndef}), 
apart from a gauge factor,  taking the form 
$ \Psi_{jk\ell} =   x^{-k}y^{-\ell}\, \psi_n$, 
where $x$ and $y$ are spectral parameters. 
By imposing this form for the wave function, together with (\ref{taured}), we find that 
(\ref{bkp})  gives a 
scalar system for the reduced wave function $\psi_n$, namely 
\beq\label{slax} 
\bear{rcl} 
\psi_{n+3}& = & -P_n\,\psi_{n+2}+x\,P_n\,\psi_{n+1}+x\,\psi_n, \\  
\psi_{n+5}& = & -x\,Q_n\,\psi_{n+3}+y\,Q_n\,\psi_{n+2}+xy\,\psi_n, \\
\psi_{n+4}& = & -R_n\,\psi_{n+3}+y\,R_n\,\psi_{n+1}+y\,\psi_n, 
\eear 
\eeq 
where we have introduced new dependent variables 
\begin{equation} \label{PQR}
P_n = \frac{1}{\delta_3}\, \frac{\uptau_{n+1}\uptau_{n+2}}{\uptau_{n}\uptau_{n+3}}, 
\quad 
Q_n = \frac{1}{\delta_1}\, \frac{\uptau_{n+2}\uptau_{n+3}}{\uptau_{n}\uptau_{n+5}}, 
\quad 
R_n = \frac{1}{\delta_2}\, \frac{\uptau_{n+1}\uptau_{n+3}}{\uptau_{n}\uptau_{n+4}}.  
\end{equation}
After identifying the prefactors from (\ref{deltas}), it is  clear that the above formulae for $P_n$ and $R_n$ are the same as   \eqref{coord_PR} rewritten in terms of tau functions. 
Using the first equation to eliminate $\psi_{n+3}$,  the third and second equations in (\ref{slax}) provide expressions 
for $\psi_{n+4}$ and $\psi_{n+5}$, respectively, as linear combinations of $\psi_{n}$, $\psi_{n+1}$ and $\psi_{n+2}$, 
that is 
\beq\label{slaxc} 
\bear{rcl}   
\psi_{n+4}& = & P_nR_n\,\psi_{n+2}-x\,P_nR_n\,\psi_{n+1}-x\,R_n\,\psi_n+y\, \phi_n^{(1)}, \\
\psi_{n+5}& = & x\, P_nQ_n\,\psi_{n+2}-x^2\,P_nQ_n\,\psi_{n+1}-x^2\,Q_n\,\psi_n+y\, \phi_n^{(2)}, 
\eear 
\eeq 
where in each case we have isolated the coefficient of $y$ as 
$$ 
 \phi_n^{(1)}=R_n\, \psi_{n+1}+\psi_n,\qquad \phi_n^{(2)}=Q_n\, \psi_{n+2}+x\,\psi_n.
$$
Now by shifting $n\to n+1$ and $n\to n+2$ in the   first equation of (\ref{slax}), we obtain alternative expressions for 
$\psi_{n+4}$ and $\psi_{n+5}$ as linear combinations of $\psi_{n}$, $\psi_{n+1}$ and $\psi_{n+2}$, 
which  combine with (\ref{slaxc}) to yield a pair of linear equations of the form 
\beq\label{linpair} 
L^{(1)}(\psi_{n}, \psi_{n+1},\psi_{n+2} )  =  y\, \phi_n^{(1)}, \quad 
L^{(2)}(\psi_{n}, \psi_{n+1},\psi_{n+2} )  =  y\, \phi_n^{(2)}. 
\eeq 
Next, setting $\phi^{(0)}_n=\psi_n$, we have $\phi^{(0)}_{n+1}=\psi_{n+1}=(\phi^{(1)}_n-\phi^{(0)}_n)/R_n$, 
and similar equations for the shifts $\phi^{(1)}_{n+1}$ and $\phi^{(2)}_{n+1}$
and so, using a tilde to denote the shift $n\to n+1$, this produces the matrix equation 
\beq \label{shiftlin} 
\tilde{\mathbf{\Phi}}={\bf M}(x)\, \mathbf{\Phi}, \qquad 
\mathbf{\Phi} =(\phi^{(0)}_n , \phi^{(1)}_n , \phi^{(2)}_n)^T,
\eeq 
where ${\bf M}$ (for $n=0$) is given by (\ref{Lax3}), including the parameter 
$\la = Q_n/( R_nR_{n+1}) =\delta_2^2/\delta_1=\delta_1\beta^2/\alpha^2$, as in (\ref{lamu}). 
To obtain the  simplest-looking version of  ${\bf M}$ we 
used 
\beq\label{mods6map} 
(\la P_n R_{n+1} -\la  R_n R_{n+1}-P_n)R_{n+2}+1=0,  
\eeq 
which is equivalent to the Somos-6 recurrence \eqref{S-6} for $\uptau_n$. 

The system (\ref{linpair}) is 
incomplete, because it lacks an equation for $y\, \phi_n^{(0)}$, but applying 
the shift $n\to n+1$ to the first equation of this pair, and using (\ref{shiftlin}), 
we obtain the missing relation. This results in the eigenvalue equation 
\beq\label{eigen} 
{\bf L}(x) \, \mathbf{\Phi} = y\, \, \mathbf{\Phi},
\eeq 
with the Lax matrix taking the form (\ref{laxm}). Upon using (\ref{mods6map}) and introducing the additional parameter 
$\mu = P_nP_{n+1}P_{n+2}/(R_nR_{n+1})=\delta_2^2/\delta_3^3=-\delta_1\beta^3/\gamma^2$, 
as in (\ref{lamu}), 
the coefficients $A_2, \dots, C_0''$ can be written in terms of $P_n,P_{n+1}, R_n,R_{n+1}$ 
and the constants $\la,\mu$, and for $n=0$ the resulting expressions coincide with those in Theorem \ref{laxp}.

Equations (\ref{shiftlin}) and (\ref{eigen}) form a linear system for $\mathbf{\Phi}$, whose 
compatibility condition is the discrete Lax equation (\ref{dlax}), 
or equivalently $\tilde{{\bf L}}={\bf M}{\bf L}{\bf M}^{-1}$, meaning that the shift $n\to n+1$ is an isospectral evolution. 
This explains the origin of 
Theorem \ref{laxp}.

\subsection*{Appendix B. Proof of Proposition \ref{SW}}

By  \eqref{big_C} and \eqref{R12}, $\bar w^2 = h(u) - v g(u) = h(0) + g(0)=0$ holds at 
${\cal R}_1, {\cal R}_2$, hence
$$
 {\tilde\pi}^{-1} ( {\cal R}_{1} ) = (u=0,v=-1,w=2,\bar w=0) , \quad
{\tilde\pi}^{-1} ( {\cal R}_2) = (u=0,v=-1,w=-2,\bar w=0) .
$$
However, observe that on $\tilde C$ the function $\bar w^2 = h(u) - g(u) v$ has zeros of order 6
at ${\cal R}_1, {\cal R}_2$. Hence $\widetilde{\widetilde C}$ is singular at the above 2 points.
To regularize it, observe that near $u=0, v=-1$ the coordinate $\bar w$ admits two Taylor expansions
\begin{align}
\bar w (u) & = \pm \mu \,(F_2+ \bar F_2) \sqrt{F_2 \bar F_2 }
\cdot u^3 + O(u^4)\, .  \label{pe_w}
\end{align}
As a result, on the regularized $\widetilde{\widetilde C}$ each of these points
${\tilde\pi}^{-1} ( {\cal R}_{1} ), {\tilde\pi}^{-1} ( {\cal R}_2)$ gives rise to a pair of points,
which we denote as $\tilde R_j^-, \tilde R_j^+$, $j=1,2$, according to the sign in \eqref{pe_w}.

Next, in view of \eqref{divs}, we have
\begin{align*}
\pi_1 \; (\tilde R_1^\pm) &= (u=0, W=(2+0)/ \sqrt{2} =\sqrt{2} , Z=0), \\
\pi_1 \; (\tilde R_2^\pm ) &= (u=0, W=(-2+0)/\sqrt{2}=-\sqrt{2} , Z=0),
\end{align*}
which, by \eqref{tt2}, gives
\begin{gather} \label{t12}
t^2(S_1^\pm ) = t^2(S_2^\pm ) =\frac{\bar F_2}{F_2}= \frac {K_2-2+\lambda-\mu}{K_2+2+\lambda-\mu} , \\
 W(S_1 ^\pm ) = \sqrt{2}, \quad W(S_2^\pm )= - \sqrt{2}. \label{Ws}
\end{gather}

To determine signs of $t(S_1^+), \dots, t(S_2^-)$, we use the expression
\begin{gather}
 t= \frac{W^2-h(u)}{4 \mu u^3 \, (1 + \lambda \,u) Q(u)\, (F_1 u + F_2) } \label{tt}
\end{gather}
obtained from \eqref{2nd}. 
However, this expression gives the indeterminate result $0/0$ for $u=0$. To
resolve it, we use the Puiseux expansions \eqref{pe_w} for $\bar w (u)$, as well as the expansion of $w(u)$,
and substitute them into $W= (w+ \bar w)/\sqrt{2}$ to get the expansions of $W^2-h(u)$ in powers of $u$.
Near $\tilde R_1^\pm$ and $\tilde R_1^\pm$ we get
$$
W^2-h(u) = \pm 2 \mu \,( F_2 + \bar F_2) \sqrt{ F_2 \bar F_2 } \cdot u^3 + O(u^4).
$$
Putting this into \eqref{tt} and taking the limit $u\to 0$,
yields 
$$
t(S_{1,2}^+) = t_* = \sqrt{\bar F_2/F_2}, \quad t(S_{1,2}^-) = - t_* \, .
$$
Finally, to find ${\cal W} (S_1^\pm), {\cal W} (S_2^\pm)$, we substitute the values of $t(S_{1,2}^\pm)$ and
$W(S_{1,2}^\pm)$ from \eqref{Ws} into \eqref{W_W}. After simplifications, we get
$$
   {\cal W} (S_1^\pm) = 4 \frac{H^{3/2}}{F_2} (1 \pm t_*), \quad
   {\cal W} (S_2^\pm) = - 4 \frac{H^{3/2}}{F_2} (1 \pm t_*),
$$
which completes the proof. 
\qed 

\end{document}